\documentclass[prd,onecolumn,showpacs,floatfix,superscriptaddress,nofootinbib]{revtex4-2}
\usepackage[utf8]{inputenc}
\usepackage{dcolumn}
\usepackage{graphicx}
\usepackage{amsmath}
\usepackage{amsfonts}
\usepackage{amssymb}
\usepackage{gensymb}
\usepackage{microtype}
\usepackage{subfigure}
\usepackage{makeidx}
\usepackage{bm}
\usepackage{epsf}
\usepackage{color}
\usepackage{multirow,dcolumn}
\usepackage{graphicx}
\usepackage{mathrsfs}
\usepackage{float}
\usepackage{cancel}
\graphicspath{{Images/}}
\usepackage{geometry}
\usepackage[colorlinks, linkcolor=blue, anchorcolor=blue, citecolor=blue]{hyperref}

\begin{document}

\title{Photon region and shadow of a rotating 5D black string} 

\author{Zi-Yu Tang}
\email{tangziyu@ucas.ac.cn}
\affiliation{School of Fundamental Physics and Mathematical Sciences, Hangzhou Institute for Advanced Study, UCAS, Hangzhou 310024, China}
\affiliation{School of Physical Sciences, University of Chinese Academy of Sciences, Beijing 100049, China}

\author{Xiao-Mei Kuang}
\email{xmeikuang@yzu.edu.cn}
\affiliation{Center for Gravitation and Cosmology, College of Physical Science and Technology, Yangzhou University, Yangzhou 225009, China}

\author{Bin Wang}
\email{wang\_b@sjtu.edu.cn}
\affiliation{School of Aeronautics and Astronautics, Shanghai Jiao Tong University, Shanghai 200240, China}
\affiliation{Center for Gravitation and Cosmology, College of Physical Science and Technology, Yangzhou University, Yangzhou 225009, China}

\author{Wei-Liang Qian}
\email{weiliang.qian@gmail.com}
\affiliation{Escola de Engenharia de Lorena, Universidade de São Paulo, 12602-810, Lorena, SP, Brazil}

\affiliation{Faculdade de Engenharia de Guaratinguetá,
Universidade Estadual Paulista, 12516-410, Guaratinguetá, SP, Brazil}

\affiliation{Center for Gravitation and Cosmology, College of Physical Science and Technology, Yangzhou University, Yangzhou 225009, China}

\vspace{3.5cm}

\begin{abstract}

To explore the possible clues for the extra dimension from the Event Horizon Telescope (EHT) observations, we study the shadow of the rotating 5D black string in General Relativity (GR). Instead of investigating the shadow in the effective 4D theory, we concern the motion of photons along the extra dimension $z$ with a conserved momentum $P_z$, which appears as an effective mass in the geodesic equations of photons. The existence of $P_z$ enlarges the photon regions and the shadow of the rotating 5D black string while it has slight impact on the distortion. The EHT observations of M87* and SgrA* can rule out the black string model with an infinite length along the extra dimension, and support the hypothesis that the extra dimension is compact to avoid the Gregory-Laflamme (GL) instability, where the length of the black string/the compact extra dimension can be constrained as $2.03125~\rm{mm} \lesssim \ell \lesssim 2.6~\rm{mm}$ and $2.28070~\rm{mm} \lesssim \ell \lesssim 2.6~\rm{mm}$ respectively.

\end{abstract}

\maketitle

\tableofcontents

\section{Introduction}

The deflection of light near a massive object has been verified since 1919, as a prediction of GR. While in the spacetime near a black hole, the gravity will be so strong that the photons can orbit the black hole and form photon spheres (regions). Such photon orbits are unstable in general, slight deviation will make the photons drop into the black hole or run away to infinity. Therefore a black hole looks like a dark disk surrounded with a shine doughnut, as shown in the black hole photos published by the EHT Collaboration \cite{EventHorizonTelescope:2019dse, EventHorizonTelescope:2022xnr}. The observation of the black hole shadow provides the more direct information of the geometry near the event horizon of the black hole, and develops as a new research focus in recent years.

On the other side, extra-dimensional theories have always been attractive and people used to pin their hope for detecting extra dimensions on high-energy experiments. After the achievement of the Gravitational Wave (GW) detection \cite{LIGOScientific:2016aoc}, physicists tried to study the features of GWs in extra-dimensional theories that can distinguish the effects of extra dimensions from those in other modified gravity theories, mainly the discrete high-frequency spectrum and shortcuts, see review \cite{Yu:2019jlb}. However, the discrete high-frequency spectrum (about $\ge 300$ GHz) is far beyond the scope of GW detectors at present, and not all GWs in extra-dimensional theories can take shortcuts. Thanks to the accomplishment of the EHT observation, the first black hole photo may provide a promising way to detect extra dimensions.

The conception of extra dimensions was first introduced by Gunnar Nordstr\"om in 1914, in order to unify electromagnetism and gravity \cite{Nordstrom:1914ejq,Nordstrom:1914fn}. Then the 5D Kaluza-Klein (KK) theory was proposed with the extra dimension to be a compact circle \cite{Kaluza:1921tu,Klein:1926tv,Klein:1926fj}, which could recover both the electromagnetism and GR in 4D spacetime. In 1983, the domain wall theory was constructed with an infinite extra dimension and a bulk scalar field \cite{Rubakov:1983bb,Rubakov:1983bz}, where the effective potential well along the extra dimension could localize the energy density of the scalar field on a 3D hypersurface, i.e. the domain wall embedded in the 5D spacetime. However, the zero mode of gravity is hard to be localized on the domain wall, and also the hierarchy problem (the huge discrepancy between the Planck scale $M_{PI} \sim 10^{19} GeV$ and the electroweak scale $M_{EW} \simeq  246 GeV$) becomes a long-standing puzzle in particle physics. Finally, Lisa Randall and Raman Sundrum (RS) proposed the well-known RS-I \cite{Randall:1999ee} and RS-II \cite{Randall:1999vf} models to solve the hierarchy problem, in which a warped structure was introduced to the compact/infinite extra dimension respectively.

Considering that the extra dimensions play an important role in the early universe \cite{Okada:1984sf}, Gregory and Laflamme made the pioneering attempt to generalize the 4D Schwarzschild black hole to 5D black string by the extension to an extra dimension with the topology  $\mathbb{S}^4_{Sch}\times \mathbb{R}^1$ \cite{Gregory:1987nb} which can be regarded as an extra hair of black holes. Compared with a hyperspherically symmetric black hole (5D Schwarzschild), a hypercylindrical black hole (5D black string) possesses higher entropy with the same mass $\mathcal{M}$ when the length of black string is small enough 
\begin{equation}
    \mathcal{S}_{BS}=\frac{4 \pi \mathcal{M}^2}{\ell}>\mathcal{S}_{Sch}=\frac{8}{3} \sqrt{\frac{2 \pi }{3}} \mathcal{M}^{3/2}~, \quad \left(\ell<\ell_0=\sqrt{27\pi \mathcal{M}/8}\right)~.
\end{equation}
In other words, an uniform 5D black string with length $\ell<\ell_0$ is thermodynamically preferred than the 5D Schwarzschild black hole, indicating a possible mechanism to trigger dimensional reduction. The 5D black string was thought to be stable under linear perturbations \cite{Gregory:1987nb}, unless the well-known GL instability of black strings/branes \cite{Gregory:1993vy,Gregory:1994bj} was addressed using the general solution of 10D black strings~branes in the low-energy string theory \cite{Horowitz:1991cd}.

A lot of efforts have been made to explore the fate of the black strings/branes instability. Gregory and Laflamme speculated that the instability could potentially fragment the horizon and form the known periodic black hole solutions (hence violating the cosmic censorship), or will not form in the first place from collapse \cite{Gregory:1993vy}. However, Horowitz and Maeda then proved that classical event horizons can not pinch off and proposed that the spacetime is most likely to settle down to a new nonuniform black string (NUBS) \cite{Horowitz:2001cz}. Almost a decade later, the numerical results of a perturbed 5D black string gave strong evidence that the classical evolution does not stop at any stable configuration but proceeds in a self-similar cascade to smaller scales \cite{Lehner:2010pn}. While for a large number of extra dimensions, the weakly NUBS was found to have larger entropy and the large $D$ approach gives a stable NUBS generically as the end point of the instability \cite{Emparan:2015gva}. Nevertheless, the fate of the black string instability with less extra dimensions is still inconclusive. (For more on instability of black strings/branes, see the review \cite{Harmark:2007md}.) 

Intriguingly, the GL instability can also be evaded by the compactification of the extra dimensions where the wavelength along the circle is required to be smaller than a critical value given by the numerical results and in agreement with the entropy argument \cite{Gregory:1993vy}. This ingenious idea was proposed initially together with the GL instability. In this work, we are devoted to explore the topology of the extra dimension (compact or not), using the EHT observations of the black hole shadow. We choose the simplest rotating 5D black strings in GR, $\mathbb{M}^4_{Kerr}\times \mathbb{R}^1$ and $\mathbb{M}^4_{Kerr}\times \mathbb{S}^1$, the latter is also a fundamental black hole solution in KK theory. In KK theory, one can start with a 4D vacuum solution of GR, then take the product with $\mathbb{S}^1$ to obtain a 5D translationally invariant solution, and finally boost the solution along the extra dimension \cite{Horne:1992zy}. When reinterpreted in 4D, such solution has nonzero charge and a nontrivial dilaton field, because the 5D vacuum GR can be dimensionally reduced to 4D Einstein-Maxwell-Dilaton (EMD) theory. Recently, the observational appearence of the most general black hole solution in KK theory has been studied elaborately in 4D, where small electric/magnetic  charges of the black holes can meet the EHT observations while they are still indistinguishable from the Kerr case \cite{Mirzaev:2022xpz}. (When the electric/magnetic charges both disappear, the general black hole solution in 5D reduces to the simple model we choose.) This result shows that the simple model $\mathbb{M}^4_{Kerr}\times \mathbb{S}^1$ can be regarded as a proper approximation for the realistic situation. As we shall see later, the reduction of the parameters is conductive to give an independent constraint for the length of the compact extra dimension, this interesting result has been briefly reported in \cite{Tang:2022hsu}.

Actually since the first black hole photo was unveiled, various attempts have been made to construct the geometry of the supermassive black hole in the center of the galaxy M87 with additional matter sources or in modified theories, including the representative extra-dimensional theories. The first quantitative constraint $\ell\lesssim 170 AU$ (here $\ell$ is the $AdS_5$ curvature radius) for the scale of the extra dimension in the RS scenario was given from the deviation of quadrupole moment from the Kerr prediction \cite{Vagnozzi:2019apd}. After that, the shadow of a rotating squashed KK black hole was studied with the specific angular momentum of photon from the fifth dimension \cite{Long:2019nox}. Later, the black hole shadow was calculated on the 4D effective brane of the 5D GR theory \cite{Banerjee:2019nnj}. In type IIB superstring/supergravity inspired spacetimes, the shadow of 5D black holes was found to be significantly distorted and shrink with the brane number \cite{Belhaj:2020okh}. Moreover, the quasinormal modes and the shadow of string-corrected $D$-dimensional black holes were investigated in \cite{Moura:2021eln}. Furthermore, the shadow of rotating braneworld black holes in the RS-II model was revisited by considering not only the metric in the near region of the black hole but also the linearized metric in the far region where the observer stays \cite{Hou:2021okc}.

This paper is organized as follows. In Sec. \ref{Sec2} we obtain the geodesic equations for a massive test particle in the rotating 5D black string spacetime, where the momentum $P_z$ along the extra dimension plays the role like an effective mass. Then in Sec. \ref{Sec3} we study the photon regions and the stability with the effects of $P_E$. In Sec. \ref{Sec4} we investigate the black string shadow observed at both finite and infinite distances, and make the parameters estimation from the EHT observations. Most importantly, in Sec. \ref{Sec5} a constraint for the length of the extra dimension can be given from the estimation of parameter $P_z/E_0$. Subsequently, in Sec. \ref{Sec6} we calculate the energy emission rate and find that existence of the extra dimension amplify the energy emission rate without changing the position of the peak. Finally in Sec. \ref{Sec7} we conclude the main results and discussions.

\section{Geodesic equations}
\label{Sec2}

In general, astrophysical black holes can be described by Kerr metric in four dimensions, if an extra spatial dimension $z$ is introduced in the simplest (uniform) way, then the constructed spacetime is still a solution of the vacuum Einstein equations of GR in five dimensions \cite{Grunau:2013oca}
\begin{equation}
    ds^2=-\frac{1}{\Sigma}\left(\Delta-a^2\sin^2{\vartheta}\right)dt^2+
    \frac{\Sigma}{\Delta}dr^2+\Sigma d\vartheta^2+\frac{1}{\Sigma}\left(\rho^4-\Delta a^2\sin^2{\vartheta}\right)\sin^2{\vartheta}d\varphi^2-\frac{4a M r}{\Sigma}\sin^2{\vartheta}dtd\varphi+dz^2~,
\end{equation}
where
\begin{eqnarray}
   &&\Sigma=r^2+a^2\cos^2{\vartheta}~,\notag \\
   &&\Delta=r^2-2Mr+a^2~,\notag\\
   &&\rho^2=r^2+a^2~.
\end{eqnarray}

This solution describes a rotating uniform black string, where $M$ is the mass density proportional to the mass of the black string $\mathcal{M}=M\ell$ ($\ell$ is the length of the black string), and $a$ is associated to the angular momentum $J=M \ell a$. When the last term in the metric disappears, the dimensions of all the quantities reduce to the normal ones in Kerr case.

If we consider the extra spatial dimension $z$ as a compact circle to avoid the GL instability, then periodic conditions along the extra dimension will be required (we assume the circumference of the compact extra dimension is the same as the length of black string $\ell$). Throughout the paper we use the units $c=G=\hbar=1$, unless the units are specifically mentioned. 

The radius of Cauchy horizon $r_-$ and event horizon $r_+$ can be obtained from  $\Delta=0$
\begin{equation}
    r_\pm=M \pm\sqrt{M^2-a^2}~,
\end{equation}
where $a^2\le a_{\text{max}}^2=M^2$. While the ring singularity is located at $r=0$ and $\vartheta=\pi/2$, solved from $\Sigma=0$. Due to the rotation, the ergoregion and causality violation region appear with $g_{tt}>0$ and $g_{\varphi\varphi}<0$ respectively, as in 4D Kerr spacetime.

The geodesic equation of a test particle (uncharged) with mass $m$ is given by 
\begin{equation}
    \frac{d^2x^\alpha}{d\lambda^2}+\Gamma^{\alpha}_{\mu\nu}\frac{dx^\mu}{d\lambda}\frac{dx^\nu}{d\lambda}=0~,
\end{equation}
where $\lambda$ is an affine parameter related to the proper time $\tau=m \lambda$. This equation can also be derived from the Lagrangian 
\begin{equation}
    \mathcal{L}=\frac{1}{2}g_{\mu\nu}\dot{x}^\mu\dot{x}^\nu~, \label{Lagrangian}
\end{equation}
where a dot over a symbol represents the derivative with respect to $\lambda$. We can calculate the momentum of the particle and the Hamiltonian
\begin{eqnarray}
   &&P_\mu=\frac{\partial \mathcal{L}}{\partial \dot{x}^\mu}=g_{\mu\nu}\dot{x}^\nu=\dot{x}_\mu~,\\
   &&\mathcal{H}=P_\mu\dot{x}^\mu-\mathcal{L}=\frac{1}{2}g_{\mu\nu}\dot{x}^\mu\dot{x}^\nu=-\frac{1}{2}m^2~.
\end{eqnarray}

To obtain the geodesic equations with separate variables, the Hamilton-Jacobi equation is required 
\begin{equation}
    \frac{\partial S}{\partial\lambda}+\frac{1}{2}g^{\mu\nu}\frac{\partial S}{\partial x^\mu}\frac{\partial S}{\partial x^\nu}=0~,
\end{equation}
where $S$ is the Jacobi action.
Considering the conservation of the rest mass $m$, energy $E_0$, angular momentum $L_\varphi$ and momentum $P_z$
\begin{eqnarray}
   &&\frac{\partial S}{\partial \lambda}=-\frac{1}{2}g^{\mu\nu}P_\mu P_\nu=\frac{1}{2}m^2~,\\
   &&\frac{\partial S}{\partial t}=\frac{\partial \mathcal{L}}{\partial \dot{t}}=-E_0~,\\
   &&\frac{\partial S}{\partial \varphi}=\frac{\partial \mathcal{L}}{\partial \dot{\varphi}}=L_\varphi ~,\\
   && \frac{\partial S}{\partial z}=\frac{\partial \mathcal{L}}{\partial \dot{z}}=P_z~,
\end{eqnarray}
a separable solution with the constants of motion $m$, $E_0$, $L_\varphi$, $P_z$ can be assumed as
\begin{equation}
    S=\frac{1}{2}m^2 \lambda-E_0 t +L_\varphi \varphi+P_z z +S_\vartheta(\vartheta) +S_r(r)~.
\end{equation}
With the Carter constant $K$ \cite{Carter:1968ks}, the geodesic equations can be finally separated as
\begin{eqnarray}
    &&\dot{t}=\frac{a^4 E_0-a^2 E_0 \sin ^2 \vartheta  \left(a^2-2 M r+r^2\right)+2 a^2 E_0 r^2+a L_\varphi \left( -2 M r\right)+E_0 r^4}{\left(a^2+r (r-2 M)\right) \left(a^2 \cos ^2 \vartheta +r^2\right)}~,\\
     &&\dot{\varphi}=\frac{L_\varphi \csc ^2(\vartheta ) \left(a^2 -2 M r+r^2\right)-a \left(a L_\varphi-2 E_0 M r\right)}{\left(a^2+r (r-2 M)\right) \left(a^2 \cos ^2 \vartheta +r^2\right)}~,\\
     &&\dot{r}=\frac{\sqrt{R}}{a^2 \cos ^2 \vartheta +r^2}~,\\
     &&\dot{\vartheta}=\frac{\sqrt{\Theta }}{a^2 \cos ^2 \vartheta +r^2}~,\\
     &&\dot{z}=P_z~,
\end{eqnarray}
where 
\begin{eqnarray}
   &&R=\left(E_0 \left(a^2+r^2\right)-a L_\varphi\right)^2-K \Delta (r)-r^2 \left(m^2+P_z^2\right) \Delta (r)~,\label{R}\\
   &&\Theta=K-a^2 \left(m^2+P_z^2\right) \cos ^2 \vartheta -\frac{\left(a E_0 \sin ^2 \vartheta -L_\varphi\right)^2}{\sin ^2 \vartheta }~.
\end{eqnarray}

In the first four equations, the momentum $P_z$ always appears as square $P_{z}^2$ together with $m^2$, indicating that the momentum along the extra dimension participates as an effective mass $\sqrt{m^2+P_z^2}$ for the particle. For photons with a momentum $P_z$ along the extra dimension, the motions along $t$, $r$, $\vartheta$, $\varphi$ directions are exactly the same as the motions of a massive particle $m=P_z$ without the extra dimension. When the extra dimension disappears or the motion of the particle is within the 4D hypersurface, i.e. $P_z=0$, the equations reduce to those for 4D Kerr spacetime, which implies that if a particle can not move along the extra dimension then we can not recognize the existence of extra dimensions via the motion of particle or shadow.

Actually the geodesic motions in this rotating black string spacetime have been comprehensively studied \cite{Grunau:2013oca}. In this paper, we focus on the photon orbits that stay on a sphere (with constant $r$), the regions accommodating such photon orbits constitute the so-called $\textit{photon region}$.

\section{Photon regions}
\label{Sec3}

For the spherical orbits, the conditions $\dot{r}=0$ and $\ddot{r}=0$ are required, which are equivalent to $R(r)=0$ and $R'(r)=0$. Using these conditions we can obtain the constants of motion $K_E \equiv K/E_{0}^2$ and $L_E\equiv L_\varphi/E_0$ selected by the radius $r_p$ of photon orbits 
\begin{eqnarray}
   &&K_E=\frac{2 r_p \Delta (r_p) }{\Delta '(r_p)^2}\left(2 \sqrt{2r_p} \sqrt{2 r_p-P_{E}^2 \Delta '(r_p)}-P_{E}^2 \Delta '(r_p)+4 r_p\right)-r_p^2 P_{E}^2~,\label{KE}\\
  &&L_E= a+\frac{r_p^2}{a}-\frac{\Delta (r_p) }{a \Delta '(r_p)}\left(\sqrt{2r_p} \sqrt{2 r_p-P_{E}^2 \Delta '(r_p)}+2 r_p\right)~,\label{LE}
\end{eqnarray}
where $P_E\equiv P_z/E_0$, the prime represents the derivative with respect to $r$ and we have set $m=0$ for photons. The condition $\Theta \ge 0$ gives the photon region
\begin{equation}
    \left(r_x \Delta_p  -\Sigma_p  \Delta_p ' \right)^2\leq 4 a^2 r_p r_x \Delta_p   \sin ^2(\vartheta )-a^2 P_{E}^2 \sin ^2(\vartheta ) \Delta_p '  \left(\Sigma_p  \Delta_p ' +2 r_p \Delta_p  \right)~,\label{PhotonRegion}
\end{equation}
where $\Delta_p \equiv \Delta (r_p)$, $\Delta_p'\equiv \Delta '(r_p)$, $\Sigma_p\equiv \Sigma (r_p)$ and
\begin{equation}
    r_x=2 r_p+\sqrt{2r_p } \sqrt{2 r_p-P_{E}^2 \Delta_p '}~.
\end{equation}
For each value of $r_p$, a range of $\vartheta$ for photon orbits can be obtained from the above inequality (\ref{PhotonRegion}). For the static case $a=0$, the photon region degenerates a photon sphere, the radius of which can be solved from
\begin{equation}
    r_x \Delta_p  -r_{p}^2  \Delta_p '=0~. \label{static}
\end{equation}

We know that the effective potential $V_{\rm eff}$ can be defined from the difference between the total energy and kinetic energy, therefore the stable condition $V_{\rm eff}''(r_p)>0$ for the photon orbits is equivalent to $(\dot{r}^2)''=(R/\Sigma^2)''<0$. Using the conditions $R(r_p)=0$ and $R'(r_p)=0$, we can obtain the stable condition $R''(r_p)<0$ with respect to radial perturbations. With the relations (\ref{KE}) (\ref{LE}), the stable condition can be expressed as
\begin{equation}
    \frac{\Delta_p'^2}{2 E_0^2 } R''(r_p)=2 r_p \Delta_p'^2 \left(2 r_p-P_{E}^2 \Delta_p'\right)-\Delta_p \left(P_{E}^2 \Delta_p'-2 \sqrt{2r_p} \sqrt{ 2 r_p-P_{E}^2 \Delta_p'}-4 r_p\right) \left(\Delta_p'-r_p \Delta_p''\right)<0~.\label{StableCon}
\end{equation}

\begin{figure}[h]
\centering%
 \includegraphics[width=.45\textwidth]{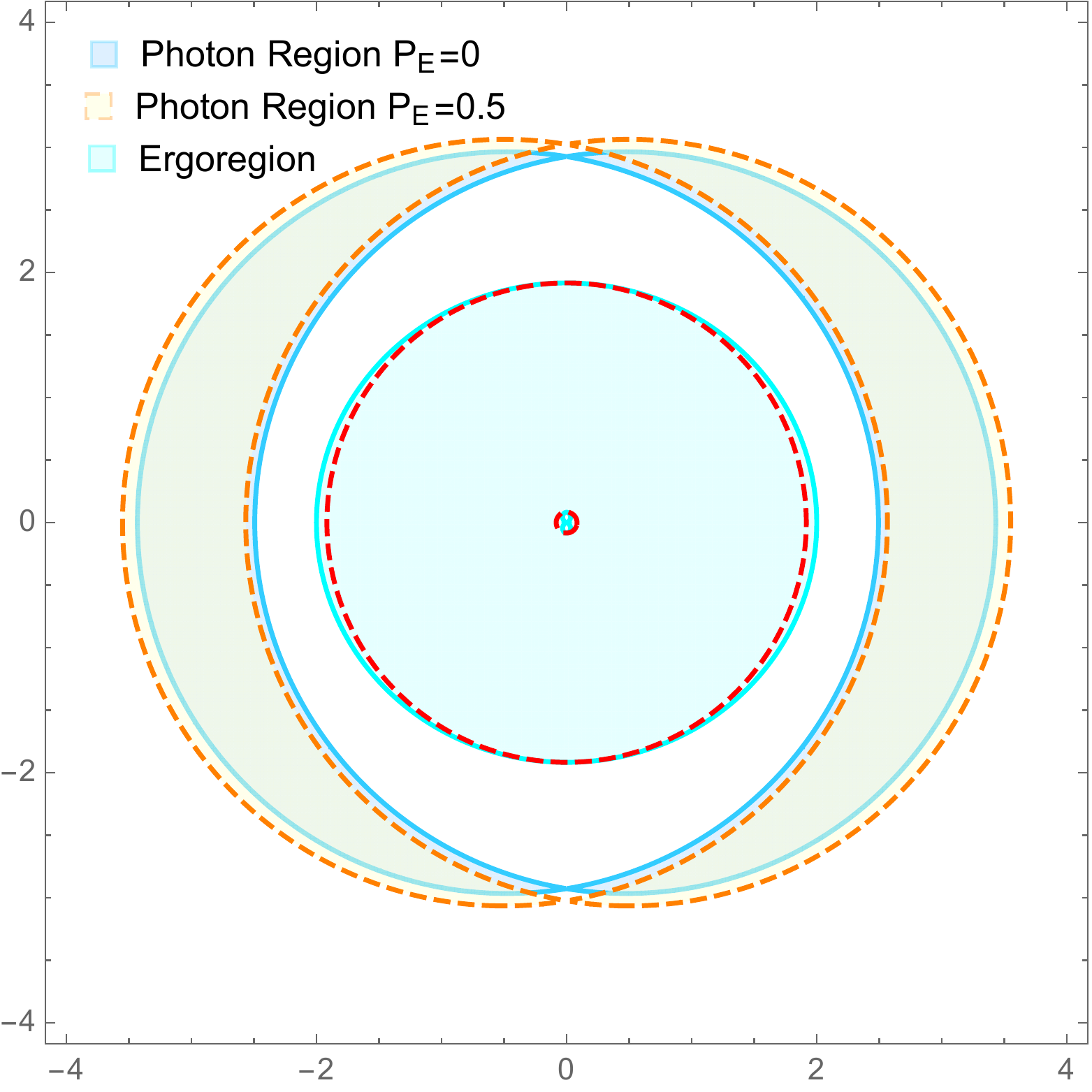} 
 \includegraphics[width=.47\textwidth]{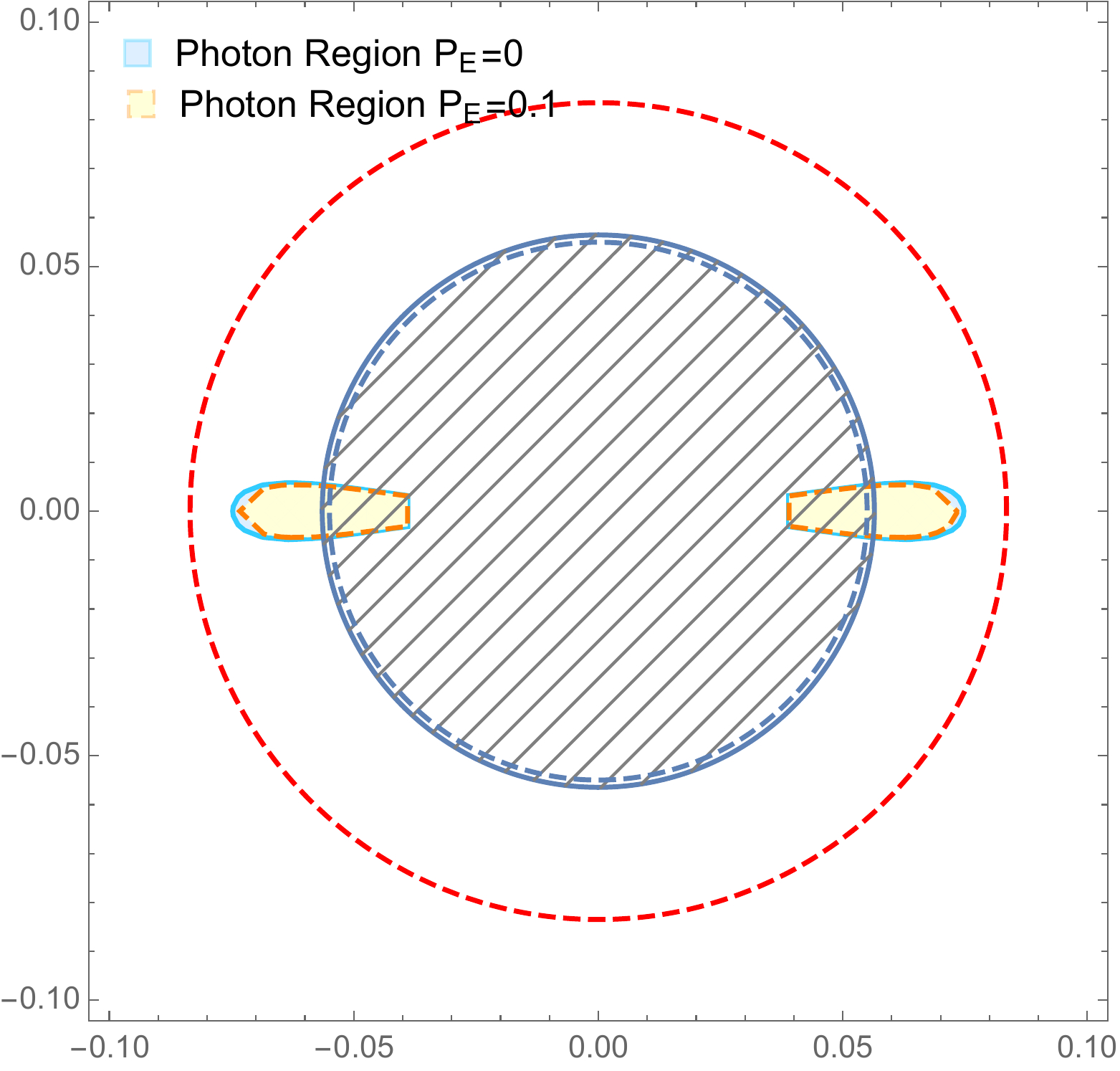} 
\caption{The photon regions outside the event horizon (left panel) and inside the Cauchy horizon (right panel) are plotted with $M=1$ and $a=\frac{2}{5}a_{\rm max}$. The parameter $P_E$ does not change the positions of horizons (red dashed circles) and ergoregion (cyan region). Besides, the stable regions (right panel) for the spherical photon orbits are filled with oblique lines, where the solid boundary is for $P_z/E_0=0$ and the dashed boundary is for $P_z/E_0=0.1$.}
\label{Fig:PhotonRegion}
\end{figure}

When $P_E \to 0$ all the relations (\ref{KE})--(\ref{StableCon}) will reduce back to those in Kerr case. In FIG. \ref{Fig:PhotonRegion}, we plot the photon regions and the stable region for the spherical photon orbits in the 5D rotating black string, compared with the Kerr case ($P_z/E_0=0$). (Although the spacetime with negative $r$ is also allowed in rotating case, for simplicity here we do not show the photon regions and the causality violation regions that with negative $r$.) It shows that with the increase of $P_E$, the photon regions outside the event horizon move out with larger radius, while the photon regions inside the Cauchy horizon move in with smaller radius. Similar with the Kerr case, the photon regions outside the event horizons are unstable while the photon regions inside the Cauchy horizons are divided into stable parts and unstable parts ($P_z/E_0=0.1$). However, if we further raise $P_z/E_0$, for example $P_z/E_0=0.5$, the photon regions inside the Cauchy horizon will disappear.

\section{Shadows of rotating 5D black string}
\label{Sec4}

The unstable photon regions outside the event horizon make the direct observation of black holes possible. The photons that escape the spherical photon orbits of a black hole due to the instability and are received by an observer in the domain of outer communication, form the boundary of the dark silhouette of the black hole. This dark silhouette is the so-called $\textit{black hole shadow}$ from the view of the observer. 

\subsection{Observer at a finite distance}

Consider an observer at position $\left(r_o, \vartheta_o\right)$ and equipped with an orthonormal tetrad
\begin{eqnarray}
   &&e_0=\frac{\rho ^2\partial_t+a \partial_\varphi}{\sqrt{\Delta  \Sigma }}\bigg|_{\left(r_o, \vartheta_o\right)}~,\\
   &&e_1=\frac{\partial_\vartheta}{\sqrt{\Sigma }}\bigg|_{\left(r_o, \vartheta_o\right)}~,\\
   &&e_2=-\frac{\partial_\varphi+ a  \sin ^2{\vartheta} \partial_t}{\sqrt{\Sigma } \sin{\vartheta}}\bigg|_{\left(r_o, \vartheta_o\right)}~,\\
   &&e_3=-\sqrt{\frac{\Delta }{\Sigma }}\partial_r\bigg|_{\left(r_o, \vartheta_o\right)}~,\\
   &&e_4=\partial_z\big|_{\left(r_o, \vartheta_o\right)}~,
\end{eqnarray}
where the timelike vector $e_0$ can be regarded as the five-velocity of the observer and the vector $e_3$ gives the spatial direction towards the center of the black hole. The tetrad has been chosen such that $e_0\pm e_3$ are tangential to the $\textit{principal null congruences}$.

\begin{figure}[h]
\centering%
 \includegraphics[width=.45\textwidth]{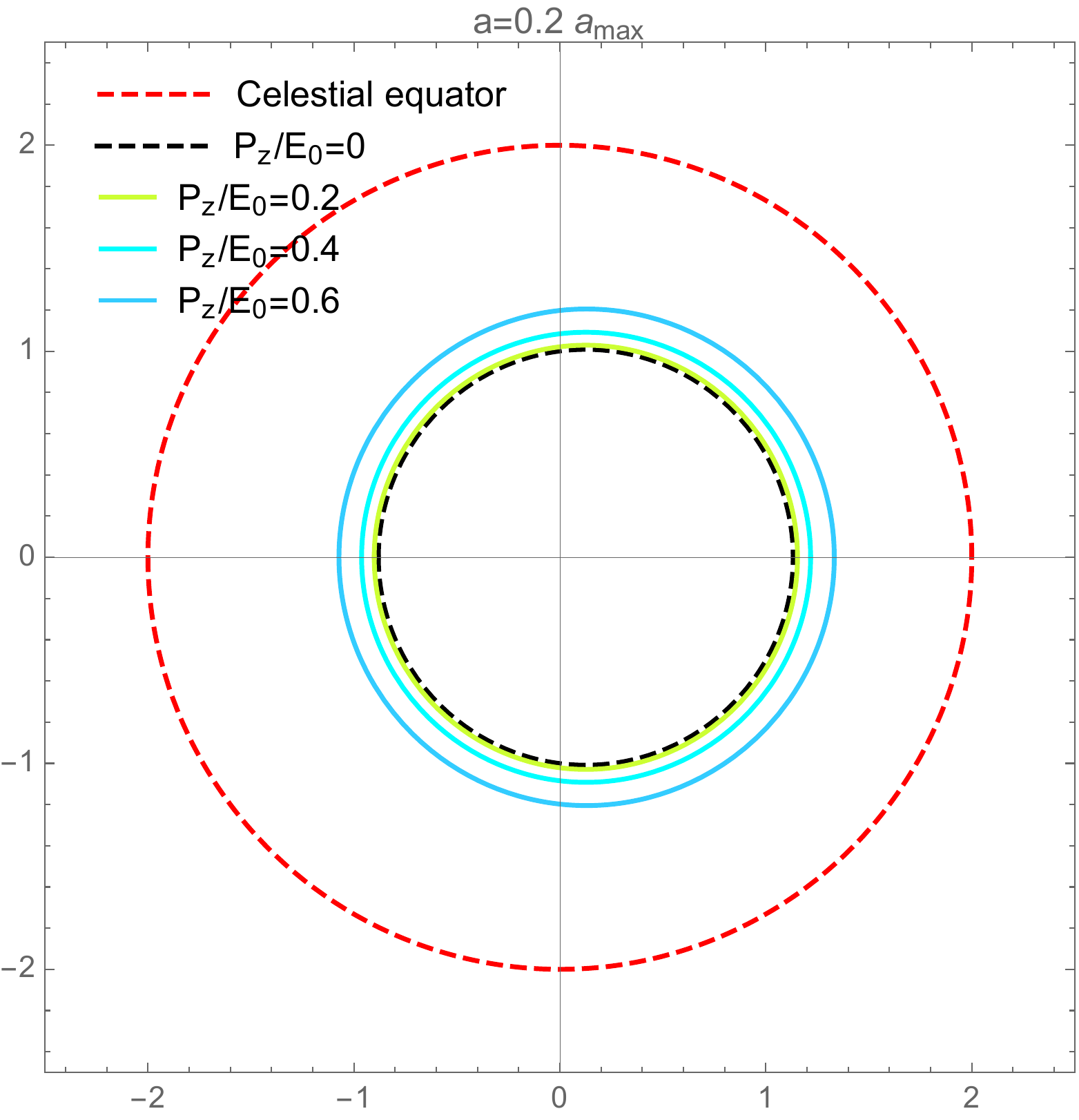} 
 \includegraphics[width=.45\textwidth]{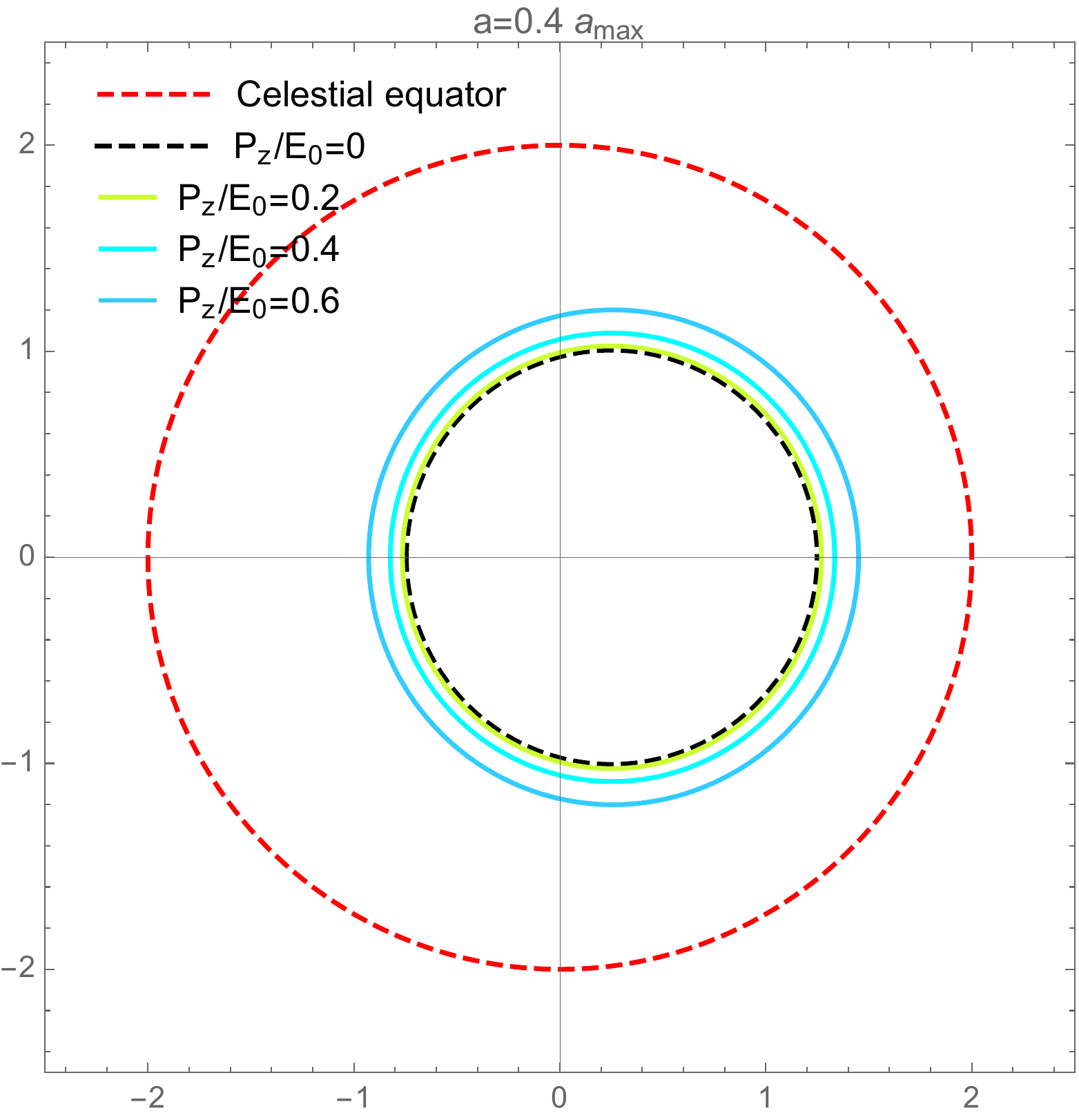} 
 \includegraphics[width=.45\textwidth]{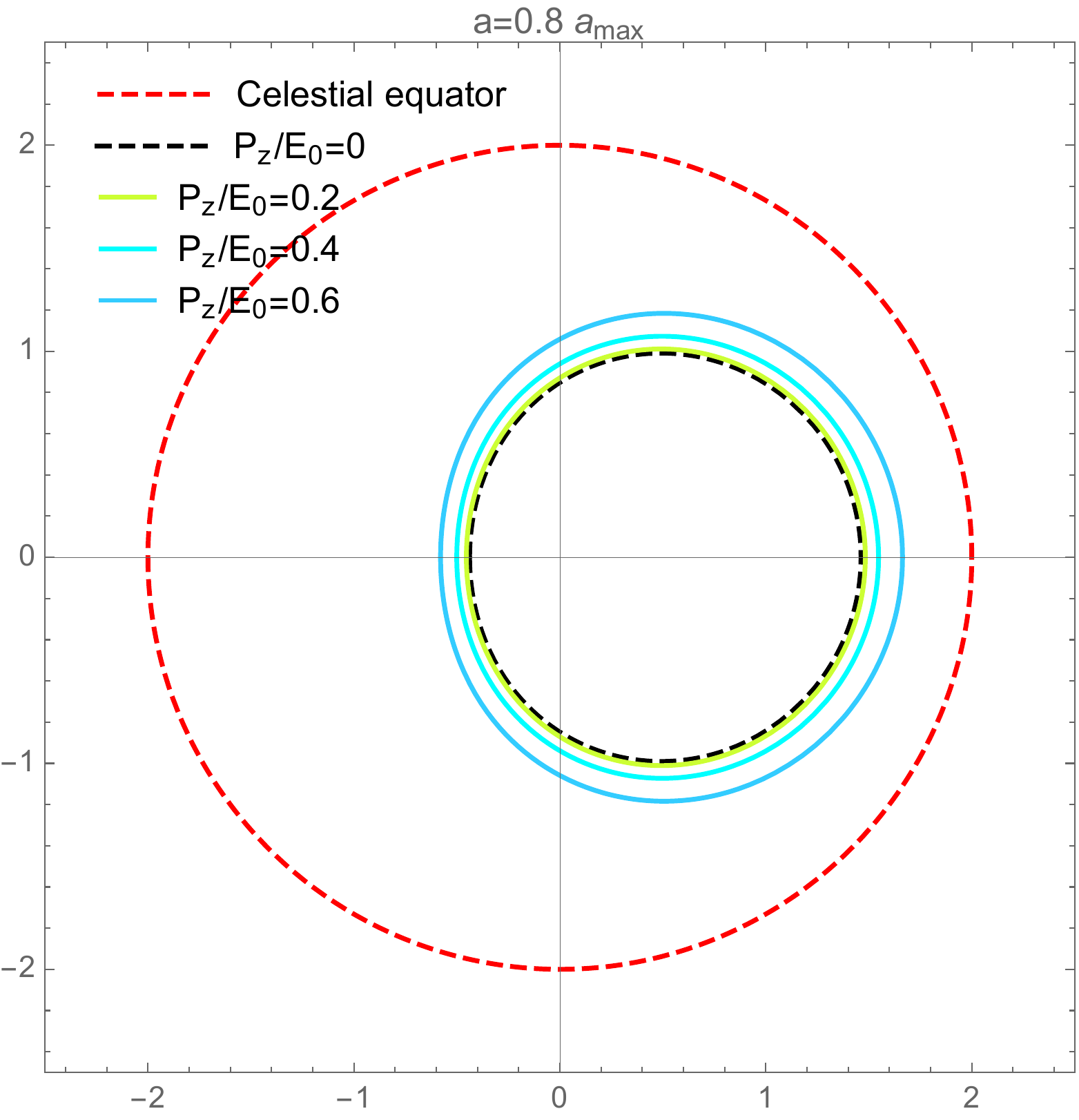} 
 \includegraphics[width=.45\textwidth]{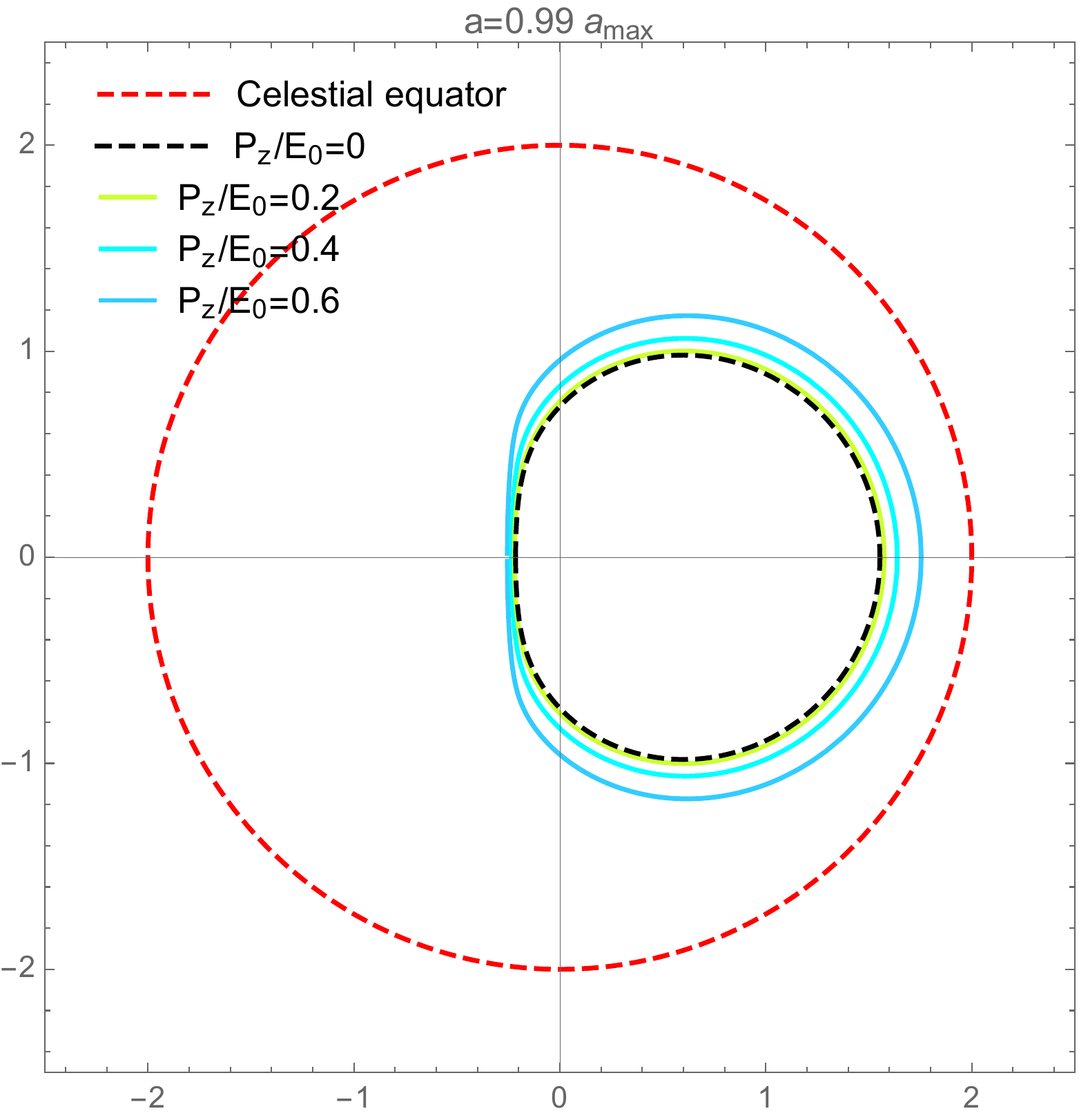} 
\caption{The shadows of 5D rotating black string being observed at $r_o=5M$ and $\vartheta_o=\frac{\pi}{2}$ with various values of $P_z/E_0$ are shown as $a=0.2~a_{\rm max}$,  $a=0.4~a_{\rm max}$, $a=0.8~a_{\rm max}$ and  $a=0.99~a_{\rm max}$ respectively. In all the figures we have set $M=1$, and the black dashed curve represents the Kerr case, while the dashed red circle is the projection of the celestial equator.}
\label{Fig:Shadow}
\end{figure}

For each light ray sent from the observer into the past, the initial direction can be described by two angles in the observer's sky (celestial coordinates), a colatitude angle $\theta$ defined as the angle deviation from the $e_3$ and an azimuthal angle $\psi$ defined as the angle deviation from the $e_1$ on the plane perpendicular to $e_3$. (The direction towards the black hole corresponds to $\theta=0$. Clear illustration can be found in FIG. 7 and FIG. 8 in \cite{Grenzebach:2014fha}.) In the black string case, an extra angle is produced as the angle deviation from the direction $e_4$ of the extra dimension. Although this angle can not be perceived by human beings, but the extra dimension also leaves imprints on the other two angles $\theta$ and $\psi$.

For fixed $\left(r_o, \vartheta_o\right)$, each value of $\psi$ corresponds to one radius $r_p$ of the spherical photon orbits that the past-oriented light ray spirals and ends on. The colatitude angle $\theta$ of this light ray is also determined by the $r_p$. A pair of such $\left(\theta,\psi\right)$ gives one point of the shadow boundary. 
For $a>0$, when the azimuthal angle $\psi$ varies from right ($\psi=-\frac{\pi}{2}$) to left ($\psi=\frac{\pi}{2}$), the corresponding $r_p$ changes from its maximal value to minimal value. This depicts the lower half of the shadow boundary and the upper half is symmetric. 

In Boyer-Lindquist coordinates, the tangent vector of the geodesic can be written as
\begin{equation}
    \partial_\lambda=\dot{t}\partial_t+\dot{r}\partial_r+\dot{\vartheta}\partial_\vartheta+\dot{\varphi}\partial_\varphi+\dot{z}\partial_z~,
\end{equation}
while at the observation event it can also be described in celestial coordinates as
\begin{equation}
    \partial_\lambda=\alpha\left(-e_0+e_1 \sin{\theta} \cos{\psi} +e_2 \sin{\theta}  \sin{\psi} +e_3 \cos{\theta} +e_4 \cos{\phi}\right)~,
\end{equation}
where the factor $\alpha$ can be obtained by comparison 
\begin{equation}
    \alpha=\frac{a L_\varphi-E_0 \rho ^2}{\sqrt{\Delta  \Sigma }}~.
\end{equation}
In this way we can assign the celestial coordinates to Boyer-Lindquist coordinates at $\left(r_o, \vartheta_o\right)$
\begin{eqnarray}
   &&\sin{\theta } =\frac{\sqrt{\Delta \left(K_E+r^2 P_E^2\right)}}{\rho^2-a L_E}\bigg|_{\left(r_o, \vartheta_o\right)}~,\\
   &&\sin{\psi }=\frac{L_E-a \sin ^2{\vartheta}}{\sin{\vartheta} \sqrt{K_E+r^2 P_E^2}}\bigg|_{\left(r_o, \vartheta_o\right)}~,\\
   &&\cos{\phi}=-\frac{P_E \sqrt{\Delta \Sigma}}{\rho^2-a L_E}\bigg|_{\left(r_o, \vartheta_o\right)}~.
\end{eqnarray}
Inserting the expressions (\ref{KE}) (\ref{LE}) for $K_E$ and $L_E$, it gives the boundary curve $\left(\theta(r_p), \psi(r_p)\right)$ of the shadow as function of $r_p$.

The shadow is always symmetric with respect to a horizontal axis, since the points $\left(\theta,\psi\right)$ and $\left(\theta,\pi-\psi\right)$ correspond to the same constants of motion $K_E$ and $L_E$. For $a>0$, the $\theta$ coordinate takes its maximal value along the boundary curve at $\psi=-\frac{\pi}{2}$ ($r_p{}_{\rm max}$) and its minimal value at $\psi=\frac{\pi}{2}$ ($r_p{}_{\rm min}$), where $r_p{}_{\rm max}$ and $r_p{}_{\rm min}$ can be solved from the equation
\begin{equation}
    \Sigma_p  \Delta_p'-r_x \Delta_p\mp a \sin{\vartheta_o} \sqrt{P_{E}^2 \left(r_o{}^2-r_p^2\right) \Delta_p'^2+2 r_p \Delta_p \left(2 r_x-P_{E}^2 \Delta_p'\right)}=0~,
\end{equation}
where $\mp$ corresponds to $\psi=\pm\frac{\pi}{2}$. 

Using the stereographic projection from the celestial sphere onto a plane (see FIG.8 in \cite{Grenzebach:2014fha}), we can express the shadow boundary in standard Cartesian coordinates as
\begin{eqnarray}
   &&x(r_p)=-2 \tan \left(\frac{\theta (r_p)}{2}\right) \sin (\psi (r_p))~,\\
   &&y(r_p)=-2 \tan \left(\frac{\theta (r_p)}{2}\right) \cos (\psi (r_p))~.
\end{eqnarray}

The shadows of 5D rotating black string being observed at $r_o=5M$ and $\vartheta_o=\frac{\pi}{2}$ with various values of $a$ and $P_z/E_0$ are depicted in FIG. \ref{Fig:Shadow}. It shows that with the increase of $P_z/E_0$, the shadow region expands larger in all the directions, while the growth of $a$ only translate and distort the shadow towards the right direction. It is known that the existence of electric/magnetic charge will make the shadow region shrinks for Kerr-Newman case, therefore the effect of the extra dimension is easily to be distinguished from the effects from other parameters like $a$ and $e$.

On the other side, extra-dimensional theories have always been attractive and people used to pin their hope for detecting extra dimensions on high-energy experiments. After the achievement of the Gravitational Wave (GW) detection \cite{LIGOScientific:2016aoc}, physicists tried to study the features of GWs in extra-dimensional theories that can distinguish the effects of extra dimensions from those in other modified gravity theories, mainly the discrete high-frequency spectrum and shortcuts, see review \cite{Yu:2019jlb}. However, the discrete high-frequency spectrum (about $\ge 300$ GHz) is far beyond the scope of GW detectors at present, and not all GWs in extra-dimensional theories can take shortcuts. Thanks to the accomplishment of the EHT observation, the first black hole photo may provide a promising way to detect extra dimensions.

Similar effects of enlargement can be found in Kerr-Newman-NUT black holes \cite{Grenzebach:2014fha} and other extra-dimensional theories. For instance, in RS scenario, a negative tidal charge enlarges the shadow of a rotating braneworld black hole and reduces its deformation with respect to Kerr spacetime \cite{Amarilla:2011fx}. It is intriguing that the shadows of black holes in most modified theories of gravity, in Loop Quantum Gravity (LQG) \cite{Brahma:2020eos} and also black holes with additional sources surrounding \cite{Afrin:2021imp} or hairy black holes \cite{Cunha:2015yba} have been studied to be smaller and more distorted compared with the Kerr black hole case. Maybe the enlargement is a representative characteristic for the black hole shadow with extra dimensions.

\subsection{Observer at infinity}

The parameter $P_E=P_z/E_0$ associated with the existence of the extra dimension is expected to be constrained from the EHT observations. For this purpose, we need to construct the shadow being observed at an infinite distance, the boundary of which can described as Cartesian coordinates on the projected plane, 
\begin{eqnarray}
   &&X=\lim_{r_o\to \infty}\left(-r_o^2 \sin{\vartheta_o}\frac{d\varphi}{dr}\right)~,\\
   &&Y=\lim_{r_o\to \infty}\left(r_o^2 \frac{d\vartheta}{dr}\right)~.
\end{eqnarray}
Substituting the geodesic equations and taking the limit $r_o\to \infty$, we can delineate the shadow boundary as
\begin{eqnarray}
   &&X=-\frac{L_E \csc{ \vartheta_o}}{\sqrt{1-P_E^2}}~,\label{X}\\
   &&Y=\pm\frac{\sqrt{-a^2 P_E^2 \cos ^2\vartheta_o-a^2 \sin ^2\vartheta_o+2 a L_E+K_E-L_E^2 \csc ^2\vartheta_o}}{\sqrt{1-P_E^2}}~. \label{Y}
\end{eqnarray}

\begin{figure}[h]
\centering%
 \includegraphics[width=.45\textwidth]{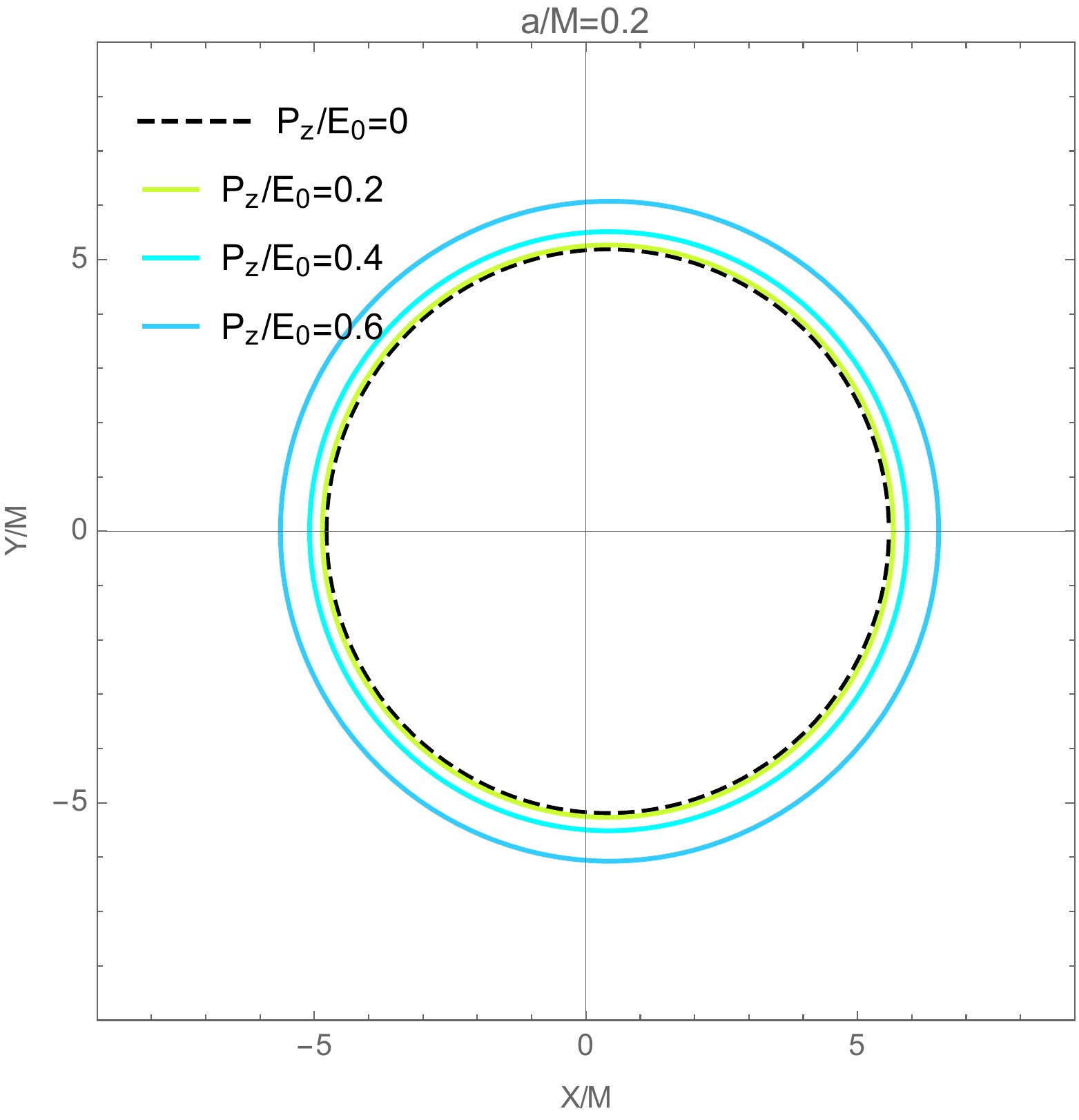} 
 \includegraphics[width=.45\textwidth]{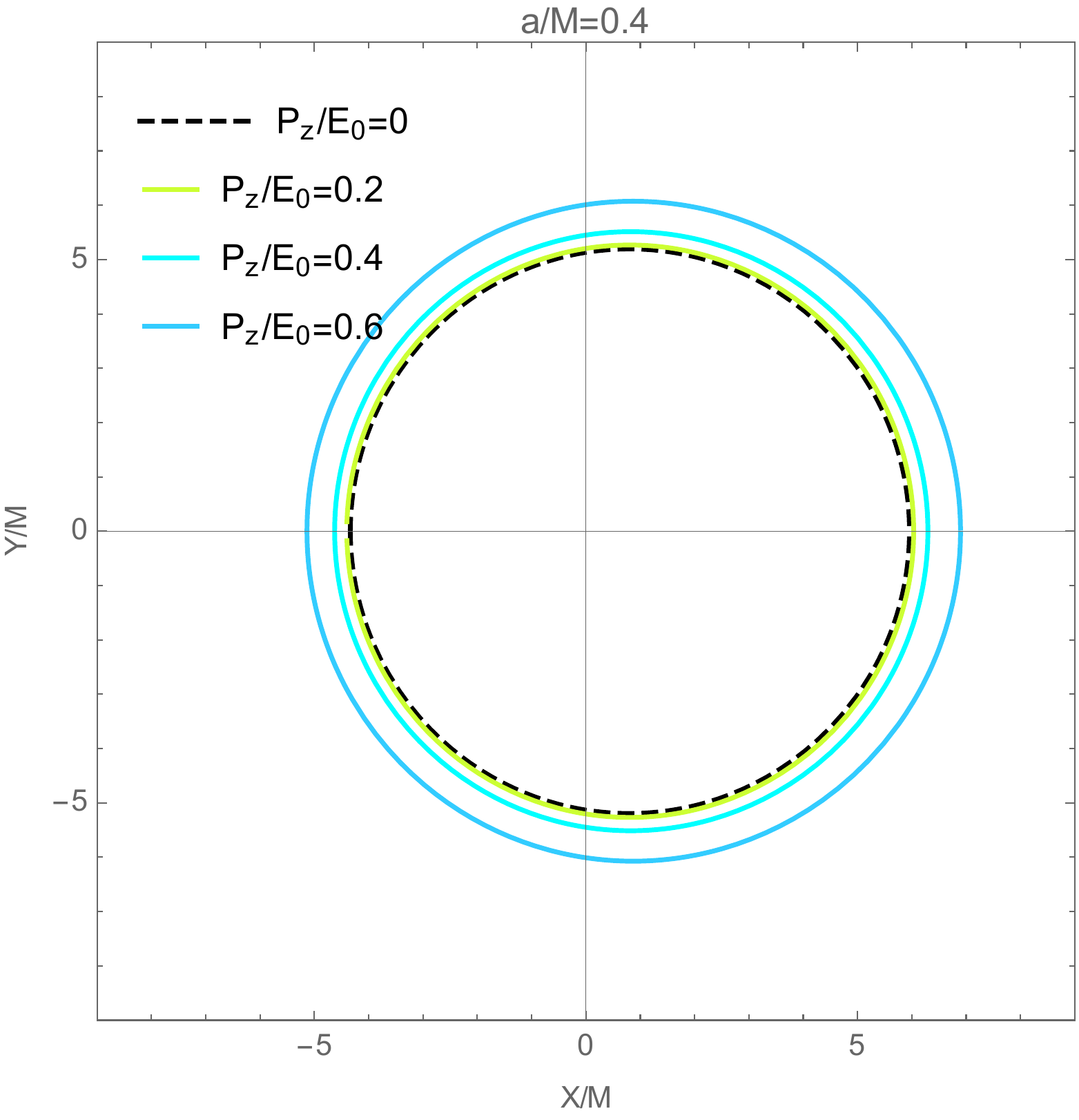} 
 \includegraphics[width=.45\textwidth]{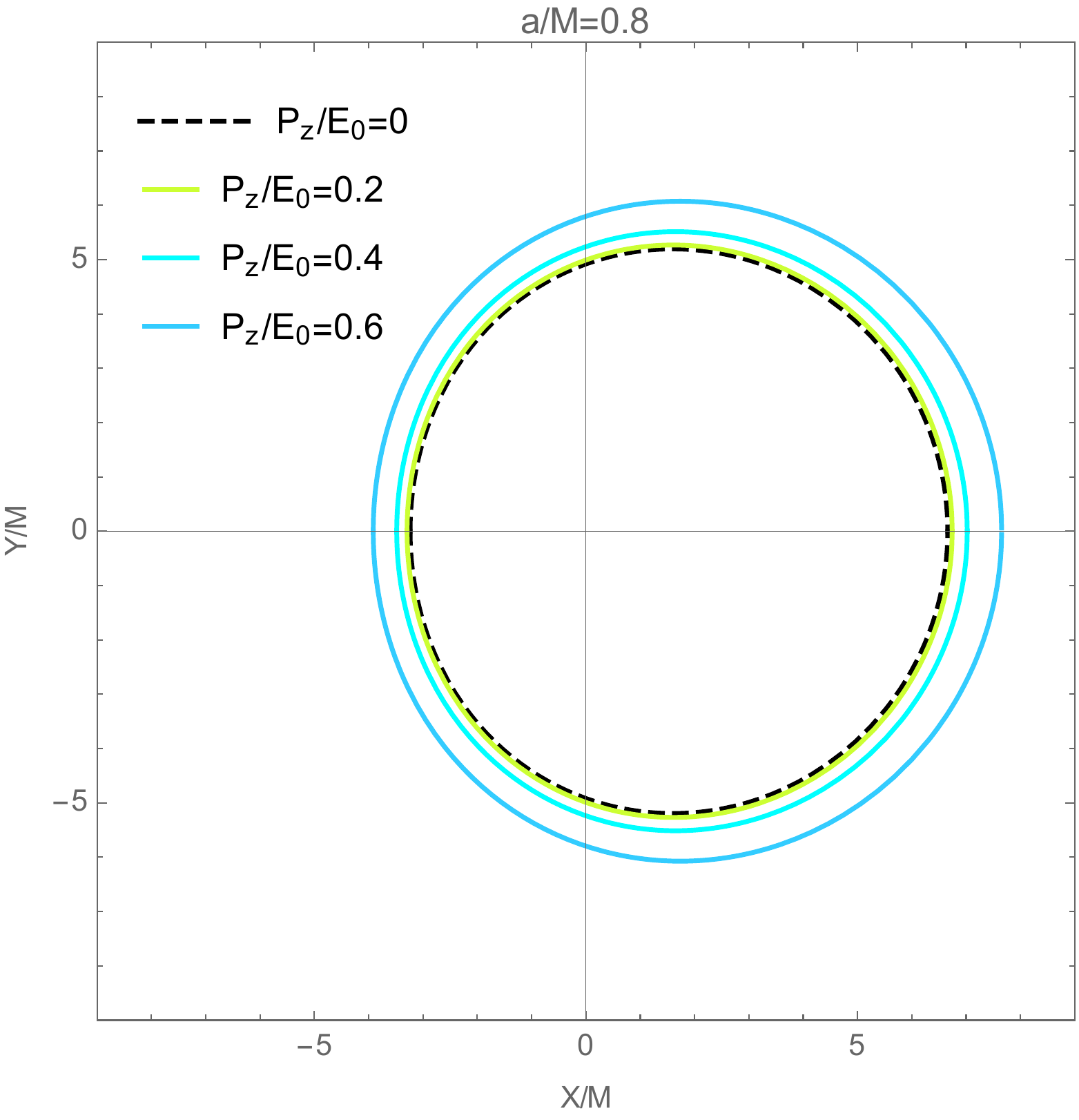} 
 \includegraphics[width=.45\textwidth]{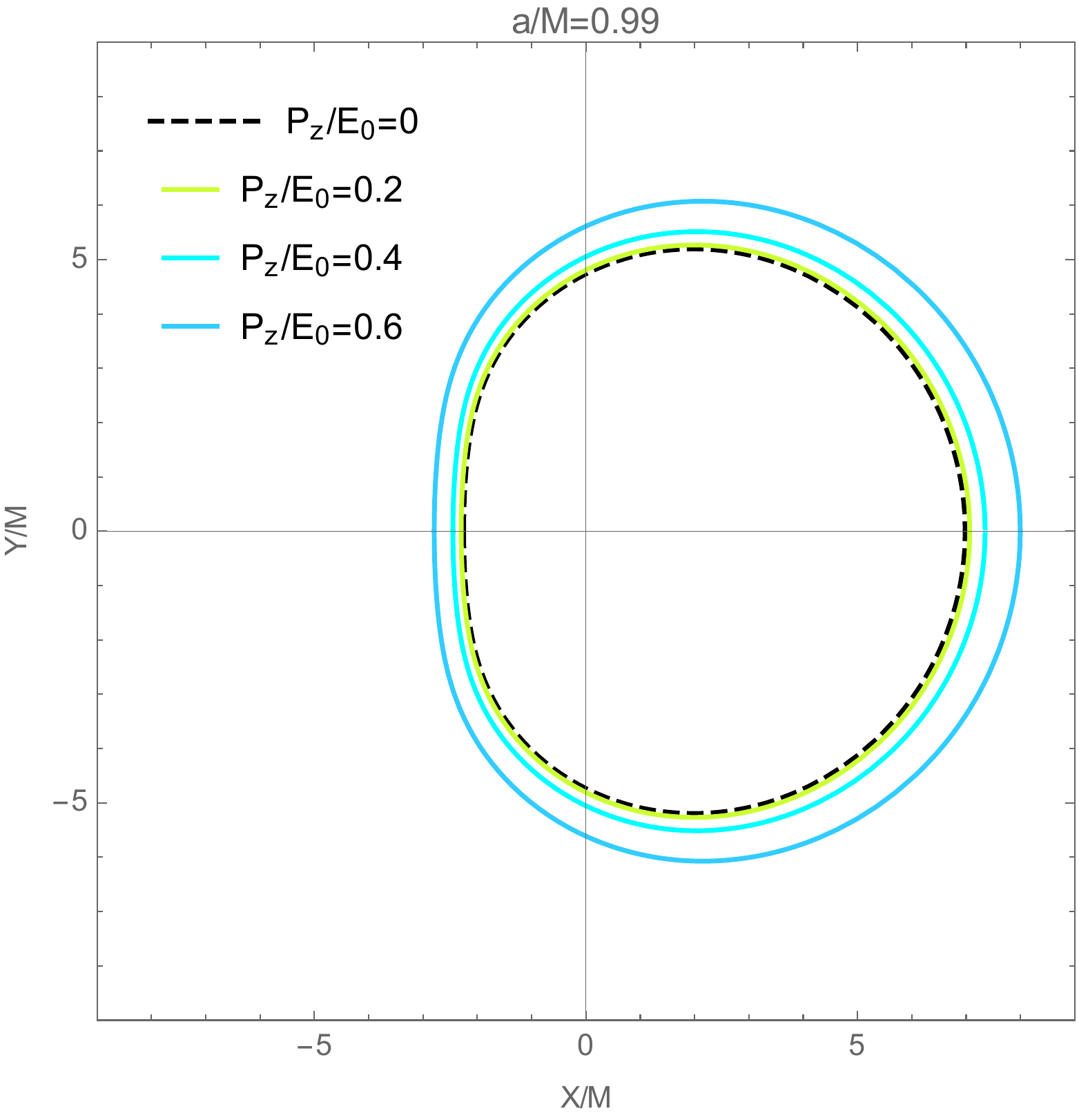} 
\caption{The shadows being observed at spatial infinity and $\vartheta_o=\frac{\pi}{2}$ with various values of $P_z/E_0$ are shown as $a/M=0.2$, $a/M=0.4$, $a/M=0.8$ and $a/M=0.99$  respectively. The black dashed curve represents the Kerr case.}
\label{Fig:ShadowXY}
\end{figure}

The shadows being observed at infinity are depicted in FIG. \ref{Fig:ShadowXY}, where the similar properties have been discussed in the previous section.

\subsection{Parameter estimation}

The boundary of shadow can be approximated by a reference circle and estimated by the radius $R_s$ of this circle and the deviation $\delta_s$ of the left edge of the shadow from the circle, proposed by Hioki and Maeda \cite{Hioki:2009na}. Using the top $\left(X_t,Y_t\right)$, right $\left(X_r,0\right)$, left $\left(X_l,0\right)$ edges of the shadow and the leftmost edge $\left(X_l',0\right)$ of the reference circle, the size and distortion of the black hole shadow can be characterized as
\begin{eqnarray}
   &&R_s=\frac{\left(X_t-X_r\right)^2+Y_t^2}{2|X_r-X_t|}~,\\
   &&\delta_s=\frac{|X_l-X_l'|}{R_s}~.
\end{eqnarray}


Later, Kumar and Ghosh defined the area $A$ and oblateness $D$ to estimate shadows with haphazard shapes
\begin{eqnarray}
   &&A=2\int Y\left(r_p\right)dX\left(r_p\right)=2\int_{r_p{}_{\rm min}}^{r_p{}_{\rm max}}\left(Y\left(r_p\right)\frac{dX\left(r_p\right)}{dr_p}  \right)dr_p~,\\
   &&D=\frac{X_r-X_l}{Y_t-Y_b}~,
\end{eqnarray}
where $Y_b=-Y_t$ in our case. For an equatorial observer in Kerr spacetime, the oblateness $D$ varies from $D=1$ (static case) to $D=\frac{\sqrt{3}}{2}$ (extremal case) \cite{Tsupko:2017rdo}.

Moreover, the average shadow radius $\bar{R}$ and the circularity deviation $\Delta C$ have been defined in \cite{Bambi:2019tjh}
\begin{eqnarray}
   &&\bar{R}=\sqrt{\frac{1}{2\pi}\int_0^{2\pi}R^2\left(\varpi\right)d\varpi}~,\\
   &&\Delta C=\frac{1}{\bar{R}}\sqrt{\frac{1}{2\pi}\int_{0}^{2\pi}\left(R(\varpi)-\bar{R}\right)^2 d\varpi}~,
\end{eqnarray}
where $R(\varpi)$ and $\varpi$ are the polar coordinates describing the shadow boundary
\begin{eqnarray}
   &&R(\varpi)=\sqrt{(X-X_c)^2+(Y-Y_c)^2}~,\\
   &&\varpi=\tan^{-1} \left(\frac{Y-Yc}{X-X_c}\right)~,
\end{eqnarray}
and $\left(X_c,Y_c\right)$ is the origin of the polar coordinates with $X_c=\left(X_l+X_r\right)/2$ and $Y_c=0$. Besides, the axial ratio $D_x$ is another way to describe the circular asymmetry by
\begin{equation}
    D_x=\frac{\Delta Y}{\Delta X}=\frac{Y_t-Y_b}{X_r-X_l}=\frac{1}{D}~.
\end{equation}

Most importantly, the angular diameter $\theta_d$ of the shadow can be obtained as \cite{Kumar:2020owy}
\begin{equation}
    \theta_d=\frac{2R_a}{d}=\frac{2R_a/M}{d/M}~,\quad R_a=\sqrt{A/\pi}~.
\end{equation}

\begin{figure}[h]
\centering%
 \includegraphics[width=.44\textwidth]{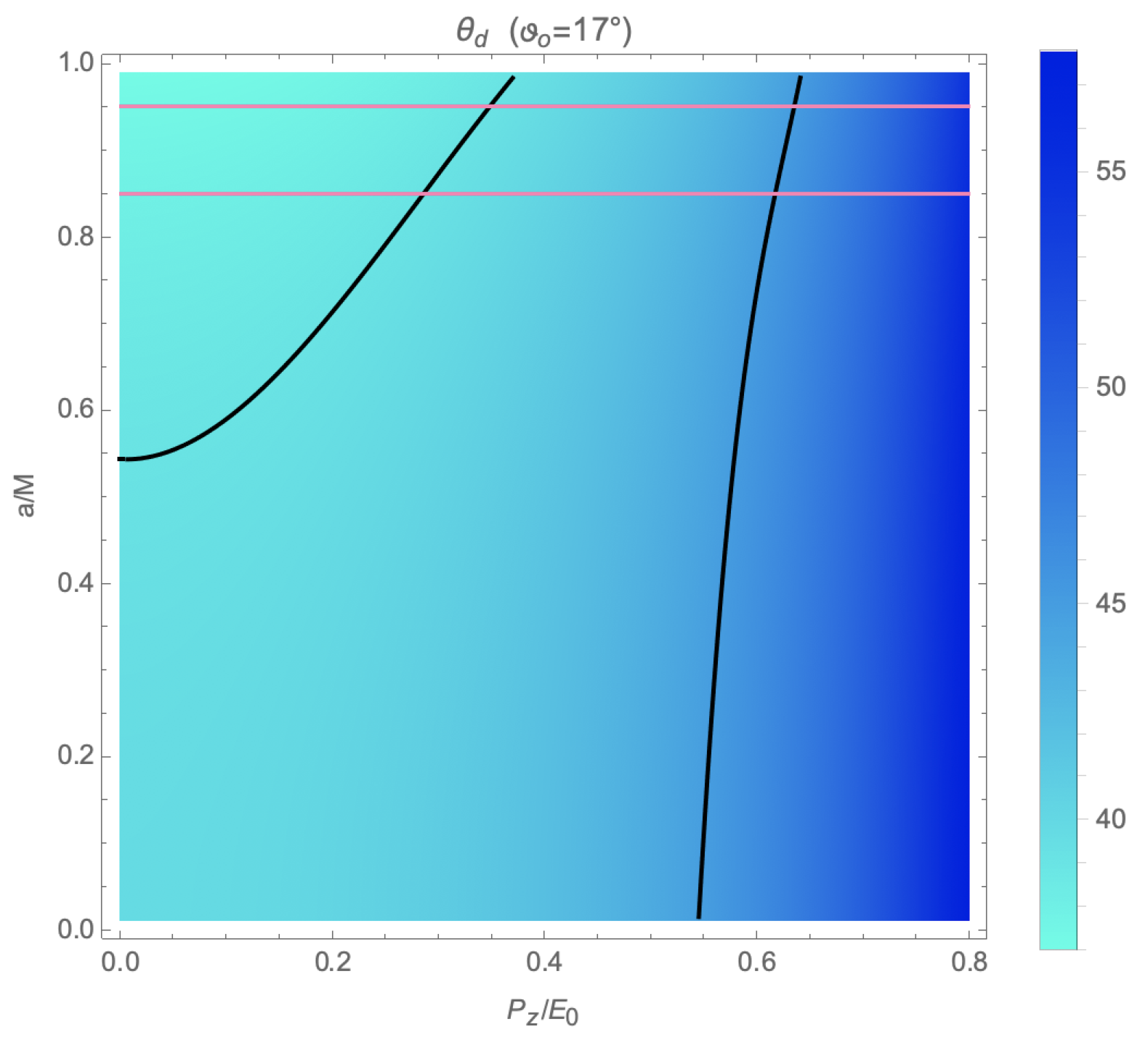} 
 \includegraphics[width=.46\textwidth]{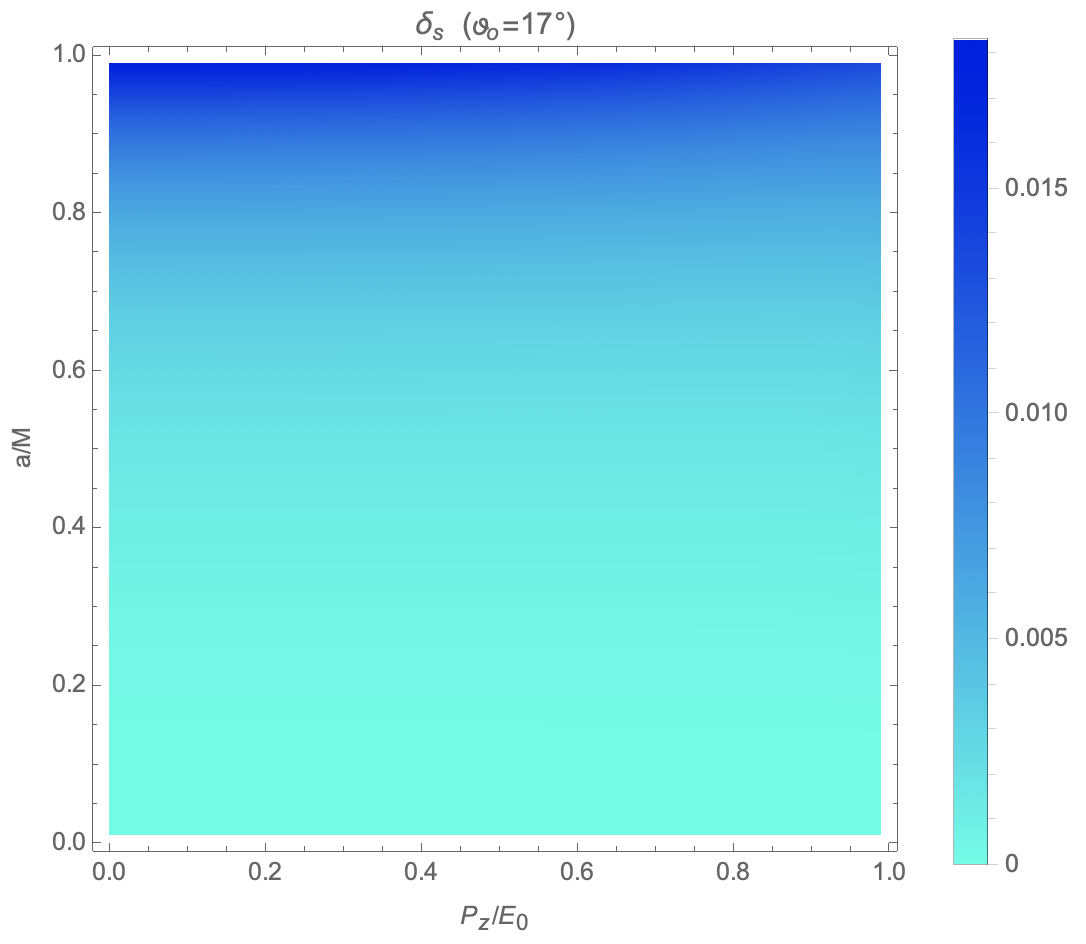} 
 \includegraphics[width=.45\textwidth]{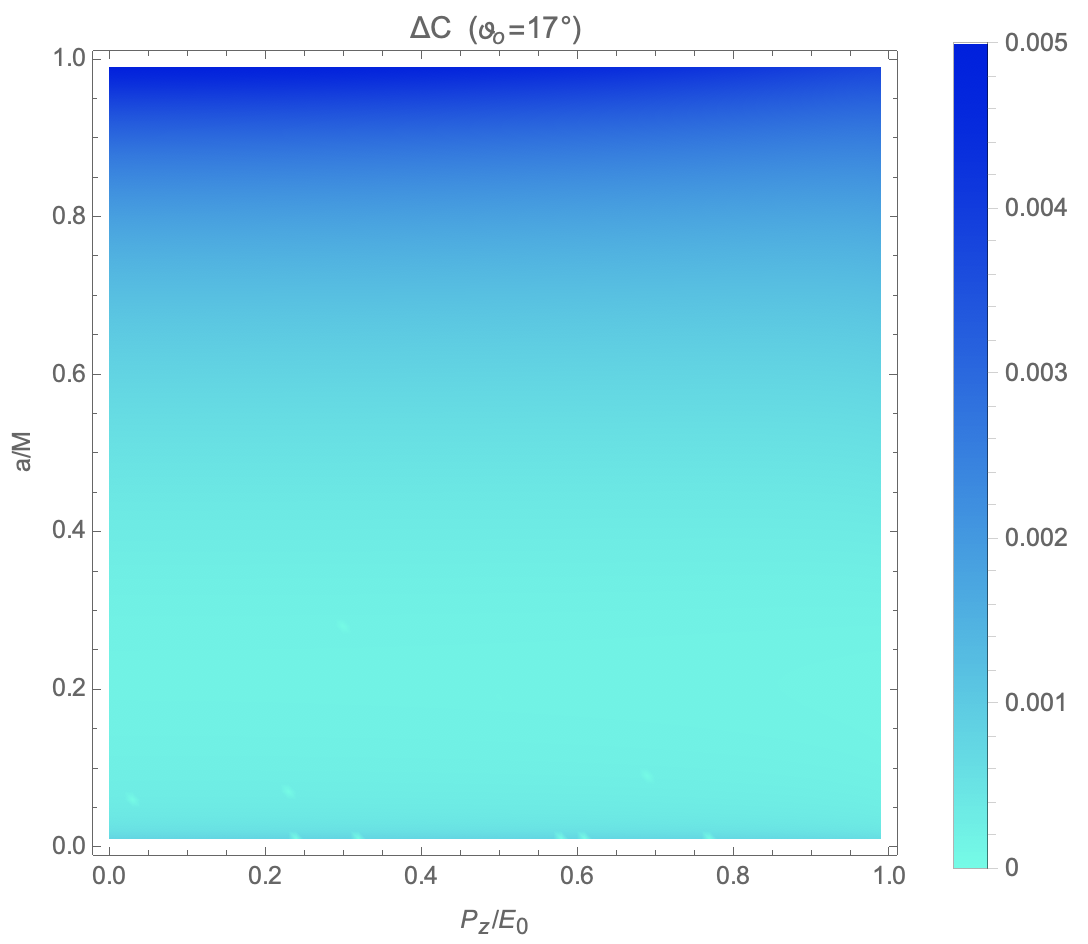} 
 \includegraphics[width=.45\textwidth]{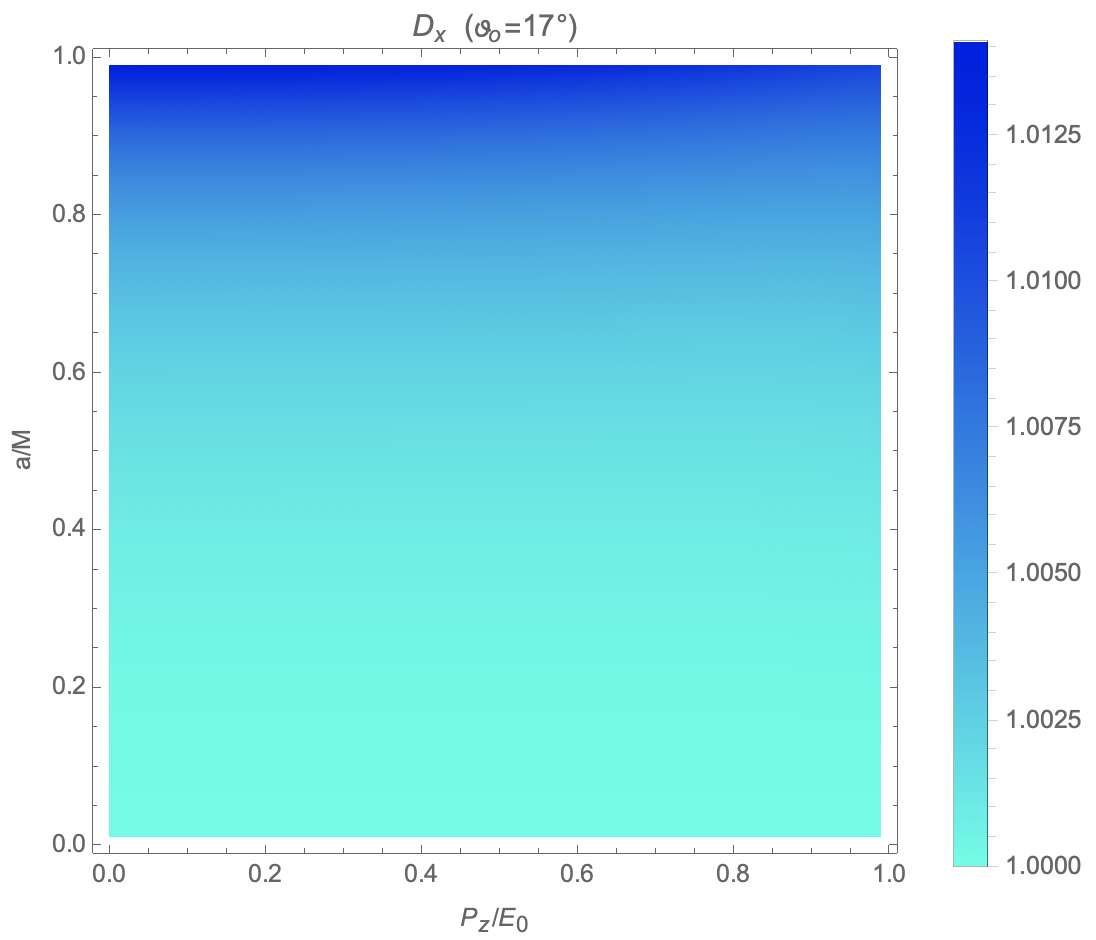}
\caption{The angular diameter $\theta_d$, deviation of the left edge of the shadow from the reference circle $\delta_s$, the circularity deviation $\Delta C$ and the axial ratio $D_x$ of the 5D rotating black string shadow are plotted as functions of $a/M$ and $P_z/E_0$, which are in agreement with the EHT observations of $M87^*$, i.e. $\Delta C \lesssim 0.1$ and $D_x \lesssim 4/3$. In the first panel, the black curves represent the ring diameter $\theta_d=42\pm 3~\rm{\mu as}$ from the EHT observations of M87*, while the two red lines denote the spin measurement $a/M=0.90\pm 0.05$ from the radio intensity data \cite{Tamburini:2019vrf}.}
\label{Fig:M87}
\end{figure}

Now we consider the realistic parameter values for the supermassive black hole in $M87^*$. Firstly the inclination angle $\vartheta_o=17\degree$ can be given by the orientation of the jets in $M87^*$\cite{CraigWalker:2018vam}. Then the distance $d=16.8\pm 0.8 Mpc$ is adopted based on the three recent stellar population measurements \cite{EventHorizonTelescope:2019ggy}. However, the mass estimates from stellar and gas dynamics does not agree with each other, here we use the average value $M=(6.5\pm 0.7 )\times 10^9 M_\odot$ inferred from the geometric models, General Relativistic Magnetohydrodynamic (GRMHD) models and image domain ring extraction, where the distance $d=16.8\pm 0.8 Mpc$ has been applied. In fact, only a ratio $d/M$ is required for the calculations
\begin{equation}
    d/M=\frac{16.8 Mpc}{6.5 \times 10^9 M_\odot}\simeq 5.40573 \times 10^{10} \quad (c=G=1)~.
\end{equation}
It is worth mentioning that here we assign the value of mass to $M$ instead of $\mathcal{M}=M \ell$, because this value is inferred from the Kerr case, in which $M$ is the mass. Using these realistic parameter values, we plot the angular diameter $\theta_d$ and other observables $\delta_s$, $\Delta C$, $D_x$ in FIG. \ref{Fig:M87}. Presupposing a Kerr black hole geometry, the EHT observations confirmed $\Delta C \lesssim 0.1$ and $D_x\lesssim 4/3$, in agreement with the predictions of black string model.

\begin{figure}[h]
\begin{centering}
    \includegraphics[width=.45 \textwidth]{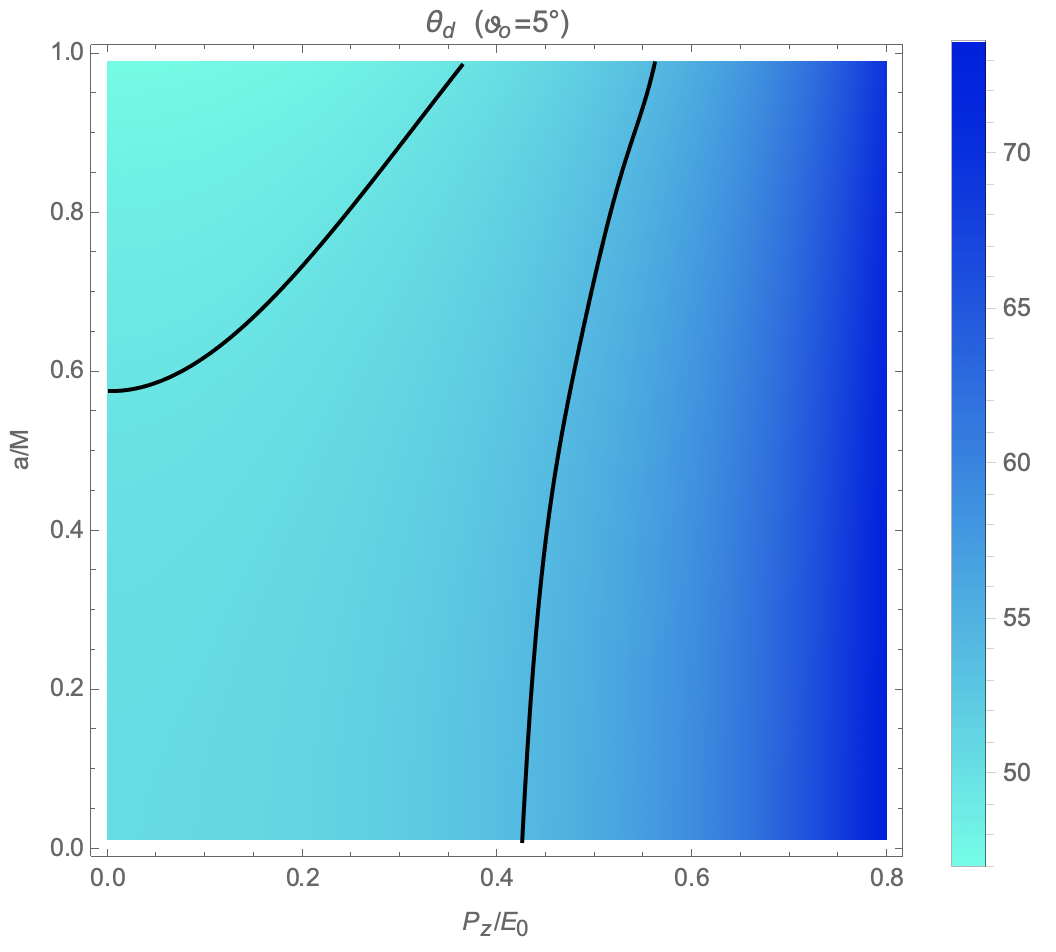}
    \par\end{centering}
    \caption{\footnotesize{The angular diameter $\theta_d$ of the 5D rotating black string shadow is plotted as a function of $a/M$ and $P_z/E_0$, compared with the ring diameter $\theta_d=51.8\pm 2.3~\rm{\mu as}$ (black solid curves) estimated from the EHT observations of SgrA*.}}
    \label{Fig:SgrA}
\end{figure}

Similarly, we also plot the angular diameter $\theta_d$ of the 5D rotating black string shadow using the EHT observations of SgrA* \cite{EventHorizonTelescope:2022xnr} in FIG. \ref{Fig:SgrA}, where $M=4.0\times 10^6 M_\odot$, $d=8121~pc$ (average distance of the three measurements) and $\vartheta_o=5^\circ$ (using the jet inclination \cite{Issaoun:2019afg}) have been applied.

\section{Constrain the extra dimension}
\label{Sec5}

Not as we anticipate before the calculations, the length of the extra dimension does not appear in the expressions of the shadow boundary (\ref{X})(\ref{Y}), but we still expect to find some clues of the extra dimension from the present results. 

Firstly, for fixed $P_z/E_0$, we can constrain the dimensionless quantity $c P_z/E_0$ from the lower bound $\theta_d=39~\rm{\mu as}$ of the EHT observation of M87* 
\begin{equation}
     \frac{P_z}{E_0}=\frac{c P_z}{E_0}=\frac{c P_z}{pc}=\frac{v_z }{c}\lesssim 0.35~,
\end{equation}
where $0.35$ is the possible maximum value of the lower bound within the spin measurement in FIG. \ref{Fig:M87}.

However, this constraint could not be proper for an infinite extra dimension, where in principle the value of $P_z/E_0$ can be arbitrary within $0\sim 1$. Especially for large enough $P_z/E_0$, the shadow boundary lies outside the outer boundary $\theta_d=45~\rm{\mu as}$ as of EHT observations, which indicates that the luminosity distribution produced by these photons is beyond the observed bright region and hence it contradicts with the EHT observations. (It is rational to suppose the luminosity distribution produced by all the photons lying right side of the right black curve in FIG. \ref{Fig:M87} is non-negligible.) 

Interestingly, a compact extra dimension can exactly give a reasonable excuse for the particular choices of $P_z/E_0$. For a compact extra dimension with a length $\ell$, the momentum $P_z$ of the photons is limited to be box normalized $P_z =2\pi\hbar n/\ell~,~n=0, \pm 1, \pm 2,~...~,$ in this way we can relate the length of the extra dimension to the momentum $P_z$ as 
\begin{equation}
    \frac{v_z}{c}=\frac{c P_z}{E_0}=\frac{2\pi \hbar n c}{E_0 \ell}=\frac{n \lambda_0}{\ell}~, \label{wavelength}
\end{equation}
where $\lambda_0$ is the wavelength of the photons. Note that for a given length $\ell$ of the extra dimension, there are infinite  choices of $n$ (i.e. choices of $P_z$), but an upper limit for the velocity $v_z/c \le 1$ can result in truncation of the larger values of $n$. 

Now let us go through all the possibilities. Primarily for $n=0$, the ground state $P_z=0$ (Kerr case) is within the range of the observation. Subsequently if the choice $n=1$ is also permitted by the observation, then the shadow boundary for the corresponding $P_z/E_0$ should be smaller than the outer boundary of the bright annulus 
\begin{equation}
    \frac{P_z}{E_0}=\frac{\lambda_0}{\ell} \lesssim 0.64~, \quad \Rightarrow \quad \ell \gtrsim \frac{\lambda_0 }{0.64} \simeq 2.03125~\rm{mm},
\end{equation}
here we have applied the EHT observing wavelength $\lambda_0=1.3~\rm{mm}$.
If the series requires to be cut off at $n=2$, i.e. 
\begin{equation}
    \frac{v_z}{c}=\frac{c P_z}{E_0}=\frac{2 \lambda_0}{\ell} > 1~, \quad \Rightarrow \quad \ell < 2\lambda_0 \simeq 2.6~\rm{mm},
\end{equation}
then the only allowed values of $P_z$ are $P_z=0$ and $P_z=2\pi \hbar /\ell$, in agreement with the observations. Finally if only choices $n=0,1,2$ are permitted, similar calculations will lead to invalid scope $4.0625~\rm{mm} \lesssim \ell \lesssim 3.9~\rm{mm}$.

For the general case, if the permitted choices are $n=0,1,2,...,k$, then we have 
\begin{equation}
    k \lambda_0/0.64 \lesssim \ell < (k+1) \lambda_0~,
\end{equation}
which is only valid for $k\le 1$. Therefore the only possible cases are $n=0$ and $n=0,1$. The former case implies the photons can not move along the extra dimension, and requires the length of the compact extra dimension to be smaller than the wavelength of the photons. While the latter case in turn constrains the length of the extra dimension as
\begin{equation}
    2.03125~\rm{mm} \lesssim \ell \lesssim 2.6~\rm{mm}~,
\end{equation}
and the parallel analysis from the EHT observations of SgrA* gives a tighter constraint 
\begin{equation}
    2.28070~\rm{mm} \lesssim \ell \lesssim 2.6~\rm{mm}~.
\end{equation}

The above analysis indicates that we could give a possible constraint on the length of the extra dimension from the EHT observations, which is surprising for the first consideration since it supports the hypothesis that the extra dimension is compact avoiding the GL instability. Nevertheless, the possibility that there is no extra dimension or the photons can not move in the extra spatial direction is still hard to rule out. Besides, it is noticed that if the length of the compact extra dimension is smaller than the wavelength of the photons, then the photons can not move along the extra dimension even if the extra dimension exists. This conclusion can be inferred from the equation (\ref{wavelength}). Therefore such constraint is based on the EHT observing wavelength and only valid when the length of the compact extra dimension is larger than the EHT observing wavelength.

It is worth noting that the critical length of the 5D static black string to avoid the GL instability is $\ell_{GL}=2\pi r_+/0.88$ \cite{Harmark:2007md}, which is much larger than the constraints we obtained.

\section{Energy Emission Rate}
\label{Sec6}

The black hole shadow observed at infinity corresponds to a high energy absorption cross-section, which oscillates around a constant limiting value $\sigma_{lim}$ for a spherically symmetric black hole. This limiting constant $\sigma_{lim}$ is the same as the geometrical cross-section of photon sphere \cite{Misner:1973prb} and for rotating black holes it can also be approximated as the area of the black hole shadow \cite{Wei:2013kza}
\begin{equation}
    \sigma_{lim} \approx \pi R_s^2~.
\end{equation}

Then the energy emission rate can be calculated as 
\begin{equation}
    \frac{d^2E(\omega)}{d\omega dt}=\frac{2\pi^2 \sigma_{lim}}{e^{\omega/T}-1}\omega^3=\frac{2\pi^3 R_s^2}{e^{\omega/T}-1}\omega^3~,
\end{equation}
where $\omega$ is photon frequency and the Hawking temperature at the event horizon 
\begin{equation}
    T\equiv \frac{\partial \mathcal{S}}{\partial \mathcal{M}}=\frac{\partial (\mathcal{S}_{\rm Kerr} \ell)}{\partial (M \ell)}=\frac{\partial \mathcal{S}_{\rm Kerr} }{\partial M}=\frac{\sqrt{-a^2+M^2}}{2 \pi\left(\sqrt{-a^2+M^2}+M\right)^2+2 \pi a^2}~
\end{equation}
is the same as the Kerr case.

\begin{figure}[h]
\centering%
 \includegraphics[width=.45\textwidth]{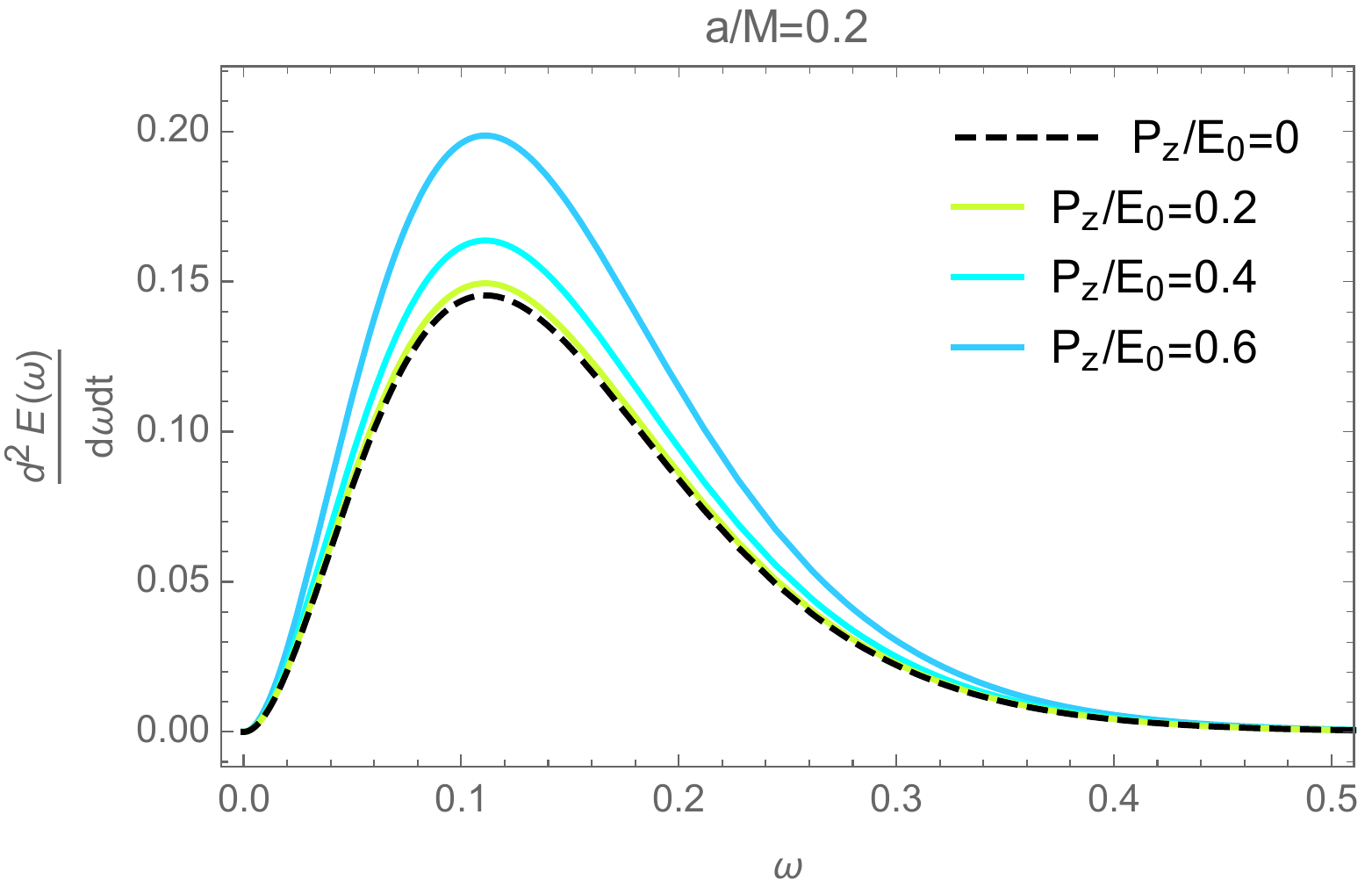} 
 \includegraphics[width=.45\textwidth]{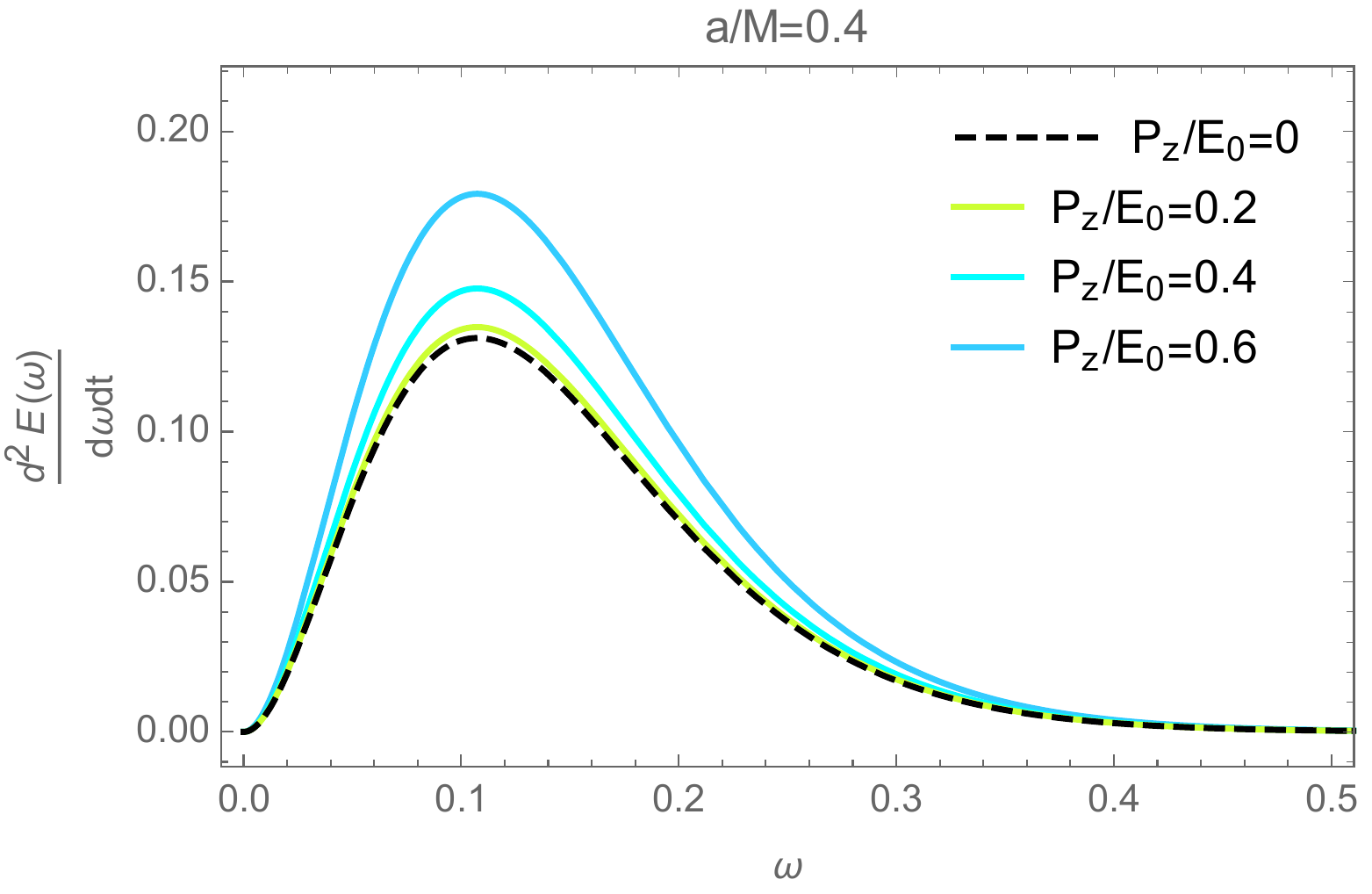} 
 \includegraphics[width=.45\textwidth]{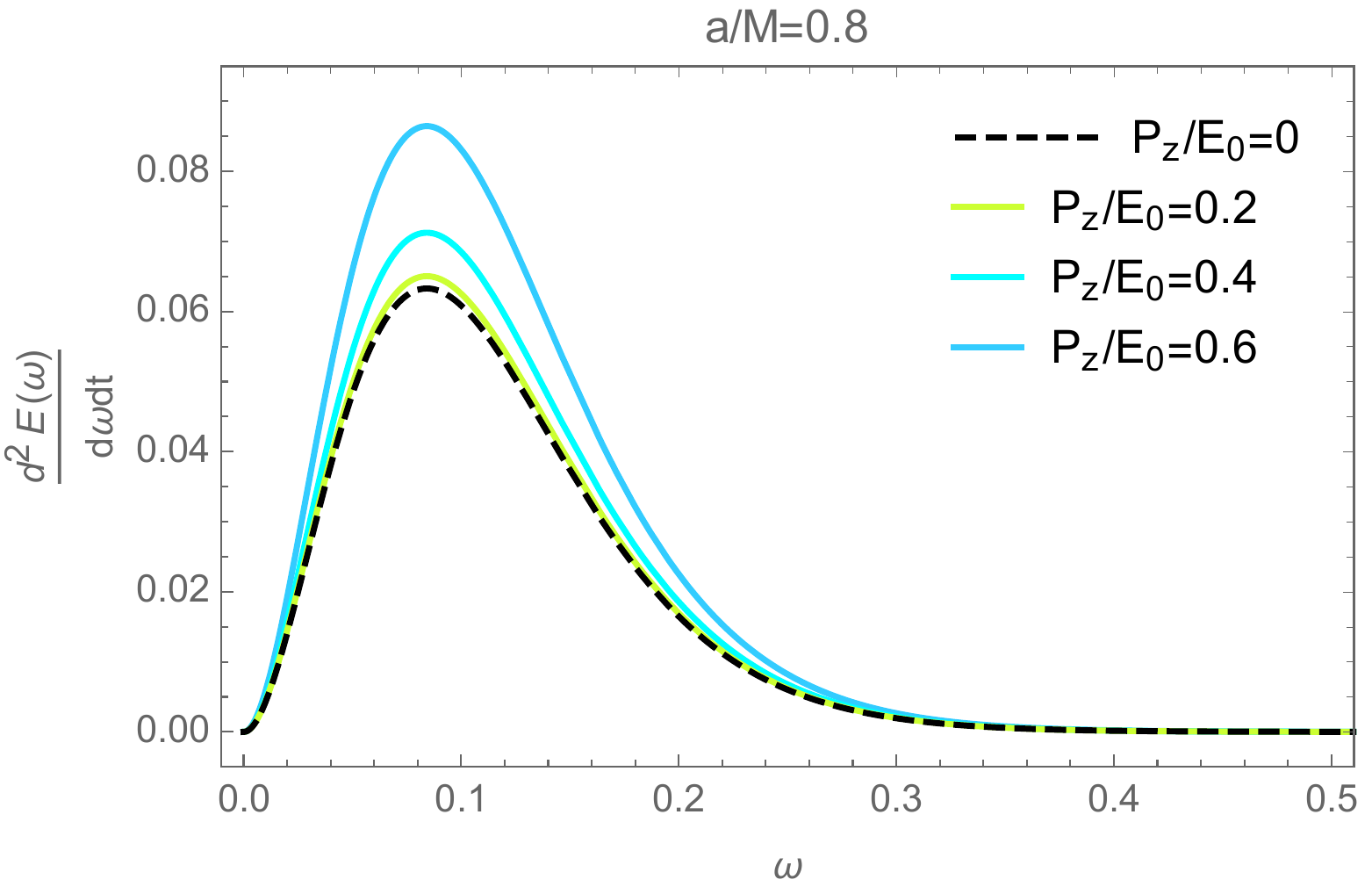} 
 \includegraphics[width=.45\textwidth]{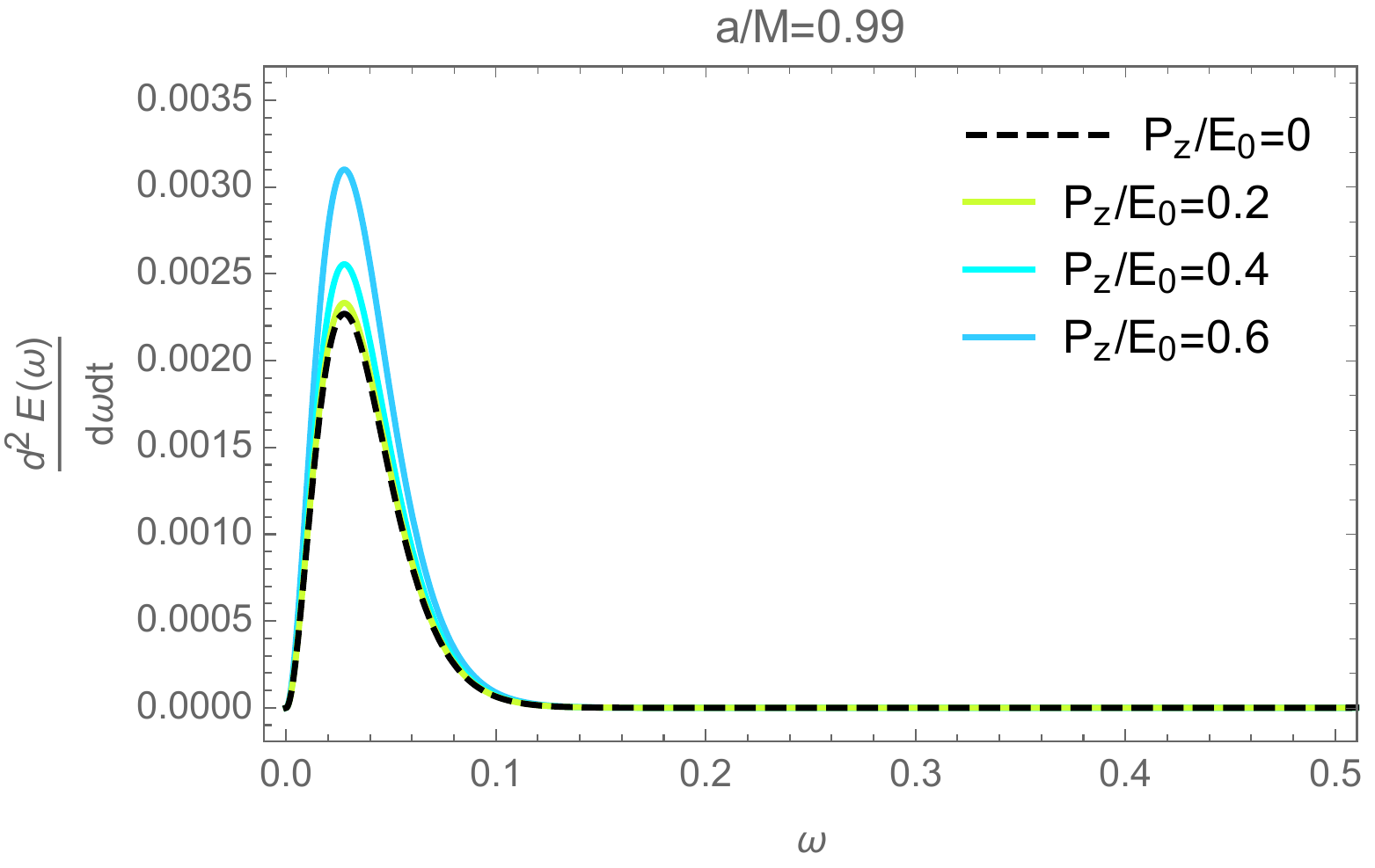} 
\caption{The energy emission rates for 5D rotating black string with various values of $P_z/E_0$ are shown as $a/M=0.2$, $a/M=0.4$, $a/M=0.8$ and $a/M=0.99$ respectively. In all the figures we have set $\vartheta_o=\frac{\pi}{2}$, and the black dashed curve represents the Kerr case. 
}
\label{Fig:Emission}
\end{figure}

In FIG. \ref{Fig:Emission}, the energy emission rate $E(\omega)$ is plotted as a function of the photon frequency $\omega$. It shows that the peak decreases and shifts to lower frequency with the increase of $a/M$. While the existence of the motion along the extra dimension makes the amplitude larger without changing the position of the peak, indicating that the energy emission will be more easily observed in the future.

\section{Conclusions}
\label{Sec7}

In this paper, we study the geodesic equations, photon regions and shadow of the 5D rotating black string in GR, with a conserved momentum $P_z$ of the photons along the extra dimension. We find that the conserved momentum $P_z$ appears as an effective mass in the geodesic equations of the photons, and enlarges the size of the shadow, while it almost has no impact on the distortion of shadow. Then we calculate various observables and compare them with the EHT observations. The results suggest that the observation requirements $\Delta C \lesssim 0.1$ and $D_x\lesssim 4/3$ from EHT can not rule out the black string in the current model.

More significantly, a constraint $cP_z/E_0 \lesssim 0.35$ can be given from the lower bound $\theta_d=39 \mu as$ of the EHT observation for M87* under the assumption that the dimensionless quantity $cP_z/E_0$ is fixed for all the photons. However, for an infinite extra dimension, there seems no proper interpretation to keep all the photons in same $P_z/E_0$ and the scenario permitting all values of $P_z/E_0$ within $0\sim 1$ is inconsistent with the observations. Intriguingly, a compact extra dimension can exactly give a reasonable excuse by the box normalization, in which the momentum $P_z$ of the photons is confined to be $P_z =2\pi\hbar n/\ell~$. In this way we can relate the length of the extra dimension to the dimensionless quantity as $c P_z/E_0=n \lambda_0/\ell$ and an upper limit for the velocity $v_z/c \le 1$ can result in truncation of the larger values of $n$. Through the careful analysis, we elicit that the only possible cases are $n=0$ and $n=0,1$, the latter case constrains the length of the extra dimension as $2.03125~\rm{mm} \lesssim \ell \lesssim 2.6~\rm{mm}$ and $2.28070~\rm{mm} \lesssim \ell \lesssim 2.6~\rm{mm}$ respectively from the observations of M87* and SgrA*.

In the end, we calculate the energy emission rate and find that existence of the extra dimension amplify the energy emission rate without changing the position of the peak. In the future work we shall study the observational appearance of the rotating 5D black string using ray-tracing method to seek a more realistic answer for the extra dimension.

\section{Acknowledgements}

We are grateful to Rong-Gen Cai, Yong-Shun Hu and Dong-Chao Zheng for beneficial discussions. This work is supported by National Natural Science Foundation of China (12147119 and 12075202), China Postdoctoral Science Foundation (2021M700142) and Natural Science Foundation of Jiangsu Province (BK20211601).

\end{document}